\documentclass[format=acmsmall, review=false, screen=true]{acmart}

\usepackage{natbib}
\usepackage{hyperref}
\usepackage{amssymb}
\usepackage{subcaption}
\usepackage{enumitem}
\usepackage{tabularx,ragged2e,booktabs}
\usepackage{rotating}
\usepackage{tikz}
\usetikzlibrary{calc,positioning,backgrounds,arrows.meta}
\usepackage{forest}
\usepackage{multirow}
\usepackage{array}
\usepackage{pifont}
\usepackage[english]{babel}
\usepackage{xspace}
\usepackage{longtable}
\usepackage{afterpage}
\usepackage[textsize=footnotesize]{todonotes}

\newcolumntype{P}{>{\raggedright\arraybackslash}p}
\newcommand{\tabitem}{~~\llap{\textbullet}~~}

\usepackage{mathtools}

\addto\extrasenglish{%

}

\newcommand\seq{sequence-aware\xspace}
\newcommand\Seq{Sequence-aware\xspace}
\newcommand\SeqA{Sequence-Aware\xspace}
\newcommand\thetitle{\SeqA{} Recommender Systems}

\newif\ifworkinprogress
\workinprogresstrue

\ifworkinprogress
 \newenvironment{revision}[1]
								{ \color{blue}  }
								{ \color{black} }
\else
 \newenvironment{revision}[1]{}{}
\fi

\newcommand{\argmax}{\mathop{\mathrm{arg\,max}}}

\usepackage[ruled]{algorithm2e} 

\SetAlFnt{\small}
\SetAlCapFnt{\small}
\SetAlCapNameFnt{\small}
\SetAlCapHSkip{0pt}
\IncMargin{-\parindent}

\setlist[enumerate]{leftmargin=*}

\acmJournal{CSUR}
\acmVolume{1}
\acmNumber{1}
\acmArticle{1}
\acmYear{2018}
\acmMonth{2}
\copyrightyear{2018}

\setcopyright{acmlicensed}

\acmDOI{0000001.0000001}

\received{July 2017}
\received[revised]{January 2018}
\received[accepted]{February 2018}

\begin{document}

\title[\thetitle]{\thetitle}

\author{Massimo Quadrana}
\affiliation{
\institution{ContentWise}
\city{Milan}
\country{Italy}
}
\author{Paolo Cremonesi}
\affiliation{
\institution{Politecnico di Milano}
\city{Milan}
\country{Italy}
}
\author{Dietmar Jannach}
\affiliation{
\institution{AAU Klagenfurt}
\city{Klagenfurt}
\country{Austria}
}
\begin{abstract}
Recommender systems are one of the most successful applications of data mining and machine learning technology in practice. 
Academic research in the field is historically often based on the matrix completion problem formulation, where for each user-item-pair only one interaction (e.g., a rating) is considered.
In many application domains, however, multiple user-item interactions of different types can be recorded
over time. And, a number of recent works have shown that this information can be used to build richer individual user models and to discover additional behavioral patterns that can be leveraged in the recommendation process.

In this work we review existing works that consider information from such sequentially-ordered user-item interaction logs in the recommendation process.
Based on this review, we propose a categorization of the corresponding recommendation tasks and goals, summarize existing algorithmic solutions, discuss methodological approaches when benchmarking what we call \emph{\seq recommender systems}, and outline open challenges in the area.


\end{abstract}


\thanks{
Author's addresses: Massimo Quadrana, ContentWise, Milan, Italy (massimo.quadrana@contentwise.com). Paolo Cremonesi, Politecnico di Milano, Dipartimento di Elettronica, Informatica e Bioingegneria (DEIB), Milan, Italy (paolo.cremonesi@polimi.it). Dietmar Jannach, Institute of Applied Informatics, AAU Klagenfurt, Austria (dietmar.jannach@aau.at).
This work was mainly carried out when Massimo Quadrana was a Ph.D. student at Politecnico di Milano.
}

\maketitle

\renewcommand{\shortauthors}{M. Quadrana et al.}

\section{Introduction}
Recommender Systems (RS) are software applications that support users in finding items of interest within larger collections of objects, often in a personalized way. Today, such systems are used in a variety of application domains, including for example e-commerce or media streaming, and receiving automated recommendations of different forms has become a part of our daily online user experience.
Internally, such systems analyze the past behavior of individual users or of a user community as a whole to detect patterns in the data. On typical online sites, various \emph{types} of relevant actions of a user can be recorded, e.g., that a user views an item or makes a purchase, and several of the actions of a single user may relate to the same item. These recorded actions and the detected patterns are then used to compute recommendations that match the preference profiles of individual users.

In academic environments, the predominant problem abstraction is that of matrix completion where we are given a user-item rating matrix and the goal is to predict the missing values. This abstraction is generally well-suited to train machine learning models that aim to capture longer-term user preference profiles. The corresponding algorithms however typically implement no specific means to take the users' short-term behavior or intents into account in their recommendations; nor are they designed to use the rich information that is contained in the sequentially-ordered user interaction logs that are often available in practical applications.

In practice, however, there are many application scenarios where considering short-term user interests and longer-term sequential patterns can be central to the success of a recommender. A typical example problem setting is that of \emph{session-based recommendation} \cite{hidasi_session-based_2016,jannach15adaptation}, where no longer-term user histories are available. Instead, we have to adapt the recommendations according to the assumed short-term interests of an anonymous user. The goal in such scenarios usually is to recommend objects that match a given sequence of user actions. 

Typical algorithmic approaches in that context learn to predict the best next item from sequential user interaction logs. Considering such sequences is however not only relevant for the short-term adaptation of the recommendations. The sequential logs can also be used to derive longer-term behavior patterns, e.g., to detect \emph{interest drifts} of individual users over time \cite{moore13taste}, to identify short-term \emph{popularity trends} in the community that can be exploited by recommendation algorithms \cite{JannachLudewigLerche2017umuai,JannachSAC2017ECommerce}, or to reason about the best point in time to \emph{remind} users of certain items they have seen or purchased before \cite{Lerche2016Reminder}. Finally, there are application domains where the recommendation of one item (e.g., an accessory) only makes sense after some other object was purchased. Such weak or strict ordering constraints might correspondingly be learned from the data and considered by a \seq recommender.

Overall, \seq recommendation scenarios are highly relevant in practice and a number of relevant works were proposed in the recent past. Research in the field is however comparably scattered and no common understanding of the different facets of the problem exists.
In this survey work, we therefore (i) categorize the various scenarios of \seq recommendations approaches in the academic literature, (ii) we review the various algorithmic approaches that were proposed to extract and leverage patterns from interaction logs, and (iii) we finally discuss specific issues when benchmarking different recommendation methods. One of the major goals of the review in that context is to lay the path for more standardized and better reproducible research works in the field.

The paper is organized as follows. In Section \ref{sec:characterizing} and Section  \ref{sec:categorization} we characterize and categorize different types of \seq recommendation problems that can be found in the literature. Section \ref{sec:algorithms} reviews existing algorithms and Section \ref{sec:evaluation} discusses methodological questions regarding their evaluation and comparison.
Section \ref{sec:conclusions} finally gives a brief outlook on future research directions. 

\section{Characterizing \SeqA Recommender Systems}
\label{sec:characterizing}
\begin{figure*}[h!t]
    \centering
    \includegraphics[width=0.75\textwidth,clip]{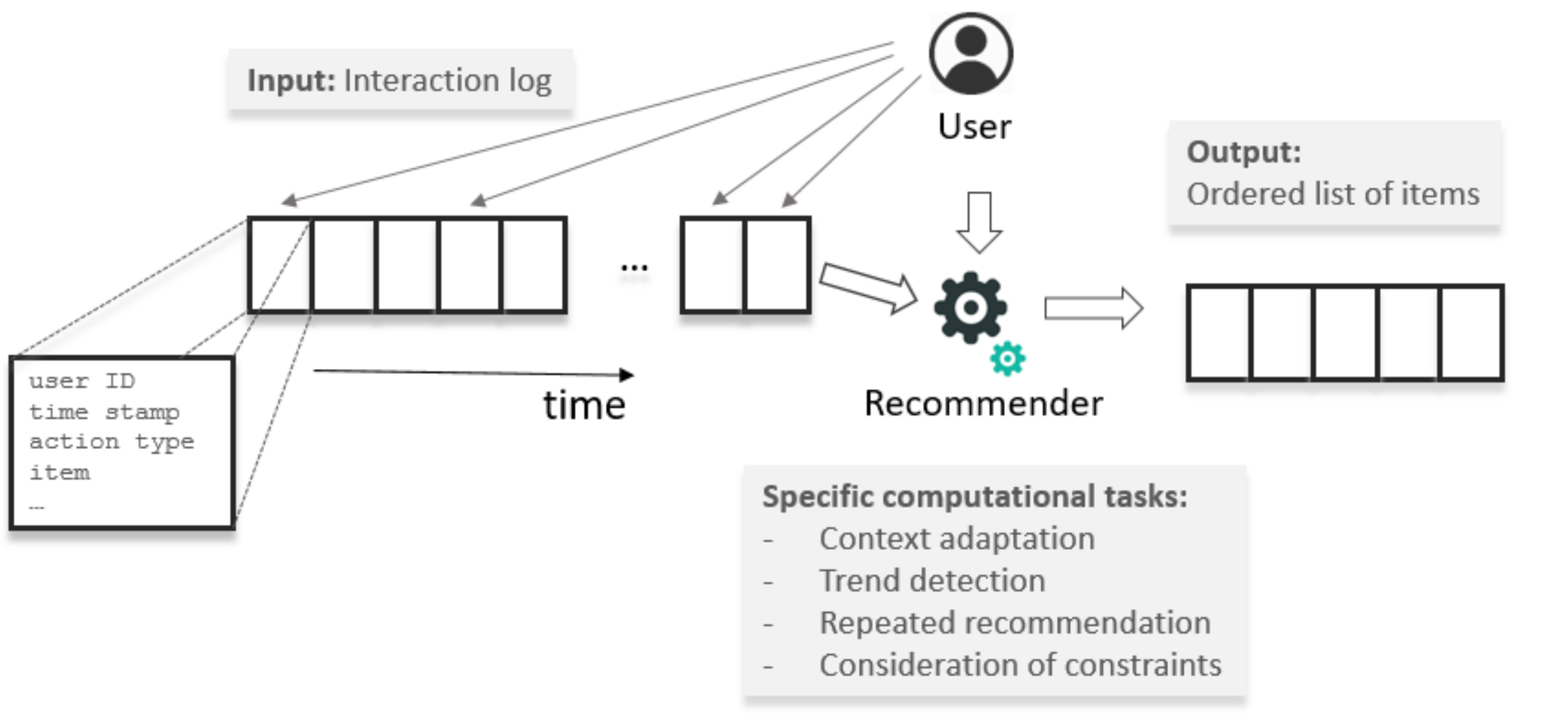}
    \caption{High-level Overview of Sequence-Aware Recommendation Problems.}
    \label{fig:problem-visualization}
\end{figure*}

\Seq recommendation problems are different from the traditional matrix-completion setup in a number of ways. Figure \ref{fig:problem-visualization} gives a high-level overview of the problem, its inputs, outputs, and specific computational tasks. In general, the ordering of the objects can be relevant 
both with respect to the inputs and to the outputs. We will discuss these aspects in more detail next.

\subsection{Inputs, Outputs, Computational Tasks, and Abstract Problem Characterization}
\paragraph{Inputs} The main input to \seq recommendation problems is an ordered and often timestamped list of past user actions. Users can be already known by the system or anonymous ones.
Each action can be associated with one of the recommendable items. Finally, each action can be of one of several pre-defined \emph{types} and each action, user, and item may have a number of additional attributes. 
Overall, the inputs can be considered as a sort of enriched clickstream data.

In the traditional matrix completion setup, all ratings are attached to one of the known users and items. We do not require this to be the case for \seq recommenders. Anonymous user actions are not uncommon, e.g., in the e-commerce domain, where users are often not logged in. Nonetheless, relevant information can be extracted from past anonymous sessions.
We also do not require that each action is related to an item, since, for example, relevant information can be extracted from the users' search or navigation behavior as well \cite{JannachLudewig2017search}.
Finally, in most application scenarios, each action will have an assigned action type (e.g., item-view, item-purchase, add-to-cart, etc.). And, depending on the domain, additional information might be available that describes further details of an action (e.g., whether an item was discounted when the action took
place), 
the users (e.g., demographics), or the items (e.g., metadata features).

Generally, such forms of input data are available in many practical applications, e.g., in the form of application or web server logs. Usually, we do however not assume to have larger quantities of \emph{explicit} ratings available in \seq recommender systems.

\paragraph{Outputs} The output of a \seq recommender are ordered lists of items as will be described more formally below. 
In this general form, the outputs are similar to those of a traditional ``item-ranking'' recommendation setup. 
However, in some \seq recommendation scenarios, the ordering of the objects in the recommendation list can be relevant as well. Instead of considering the list of recommendations as a set of \emph{alternatives} for the user, there are scenarios where the user should consider \emph{all} recommendations and do this in the provided order. Typical examples include the recommendation of a sequence of tracks in music recommendation or the recommendation of a series of learning courses. We will describe such application scenarios in more detail later in the paper.



\paragraph{Computational Tasks}
Different computational tasks of \seq{} recommenders can be identified in the literature.
Most commonly, a task that is not present in traditional matrix completion setups is the identification of sequence-related patterns in the recorded user actions. These can be \emph{sequential} patterns, where the order of the actions is relevant, or they can be \emph{co-occurrence} patterns, where it is only important that two actions happened together, e.g., within the same session. In some cases, also \emph{distance} patterns can be relevant, e.g., when the problem is to compute a good point in time to remind the user of something through a recommendation. Note that the corresponding patterns do not have to be made explicit, as often done, e.g., in sequential pattern mining \cite{Mabroukeh:2010:TSP:1824795.1824798}, but can be implicitly encoded in complex machine learning models as well. 

Besides the identification of such patterns that are subsequently used in the recommendation task, another computational task of a \seq recommender can be to reason about \emph{order constraints}. Such constraints can be either prescribed and given for an application domain as strict constraints  (e.g., in terms of a given curriculum for the learning course recommendation problem), given as heuristics (e.g., in terms of track transition rules for next-track music recommendation), or be implicitly derived from the given input data as a sort of weak constraints.

Finally, the patterns (or more generally, the learned models) that were identified in the data and the constraints have to be related with the point in time for which the recommendation is sought for. In a session-based recommender, one might consider the last few user actions and then look for past sessions that were similar to the current one. On the other hand, when a recommender is used as a reminder for the repeated purchase of consumables, the distance (in time) to the last purchase action of the user to the present time might be relevant.

\paragraph{Abstract Characterization}

Adopting the formalisms of \cite{Adomavicius:2005:TNG:1070611.1070751}, we can describe the problem at a more formal, abstract level as follows. 
Let $C$ be a set of users and $I$ a set of recommendable items.
In contrast to matrix-completion problems, we are not interested in predicting a utility value for each $i \in I$ and for each $c \in C$, but in computing an ordered list 
of objects $L$ of length $k$ for each user, where each element of $l \in L$ corresponds to an element of $i \in I$.
Technically, each sequence $L$ is an element of the set of all permutations up to length $k$ of the powerset of $I$, i.e., $L \in S_{k}(\mathcal{P}(I))$.
We denote this latter set of possible lists as $L^*$.

Let $u$ be a function that returns a utility score of a given sequence $L$ for a user $c$, 
i.e., $u: C \times L^* 
\rightarrow R$. 
The sequence-aware recommendation problem then consists of determining the sequence $l'_{c} \in L^*$ that maximizes the score for the user, i.e., 
\begin{equation}
\label{eq:optimization-problem}
\forall c \in C,\mbox{~~~}l'_{c} =  \argmax_{l \in L^*} ~u(c,l)
\end{equation}

The main problem in recommender systems is to learn or extrapolate the utility function $u$ from some given data. In the matrix completion problem, which underlies the work in \cite{Adomavicius:2005:TNG:1070611.1070751}, the input is a sparse matrix of user-item ratings.
In sequence-aware recommender systems, we in contrast assume that the underlying data is a dataset $D$ consisting of sequence\footnote{A sequence, as usual, is considered as an ordered set of objects.} of user actions where each user action $A \in D$ has a number of attributes.
A sequence dataset $D$ can be considered as an enriched log of actions of a user community, where the attributes of each action $A$ includes some sort of user ID\footnote{In practice, the user ID can either refer to a known user or it is created from a cookie in an ongoing user session.} and additional optional attributes like the \emph{action type} (e.g., an item view or click event) or a timestamp.

Overall, our function $u$ is not limited to characterizing utility scores for individual items, but for entire ordered lists of items. This makes it possible to consider additional aspects of utility in sequence-aware recommendation problems, including the diversity of the set as a whole, the quality of the ordering itself, e.g., in terms of transitions between objects, or the degree of fulfillment of weak or strict order constraints in $L$. How these quality factors are considered within existing algorithms will be discussed later in this work in Section \ref{sec:algorithms}.

Generally, the design of the utility function $u$ depends on the specific type of value the recommender system should provide to the user, or its \emph{purpose} in the sense of \cite{JannachAdomavicius2016purpose}. In the literature on recommender systems, researchers often do not explicitly discuss the underlying purpose of the system, which could be information filtering but also discovery support. Instead, they focus on optimizing an abstract computational task like predicting a hidden rating.
The situation in the context of \seq{} recommenders is often similar with the difference that the computational task is mainly to predict the hidden elements of a session given the session beginning.
While the performance of an algorithm that predicts the next hidden user action can be assessed with standard measures from information retrieval (such as precision and recall), in other cases specific measures (e.g., diversity metrics) are required in the evaluation process.



\subsection{Relation to Other Areas}
\paragraph{Implicit-Feedback Recommender Systems}
Our characterization of the \seq recommendation problem mainly targets scenarios in which we observe the individual and collective behavior of a user community over time instead of asking for explicit item ratings. A number of research works exist that focus on \emph{implicit} user feedback like purchase events. The problem formulation is however often based again on matrix completion, where multiple interactions of one user with an item are not taken into account. \emph{Explicit} item ratings, on the other hand, \emph{can} also be taken into account in a \seq recommender as one of several types of user actions.
One potential problem in that context however is that the point in time when users provide a rating can be quite different from the point in time when they consumed or purchased an item (e.g., when registering for a movie recommendation service, users initially rate a bunch of movies they have watched in the past). 
The sequence and timestamp of the ratings might therefore mislead a \seq recommender.


\paragraph{Context-Aware and Time-Aware Recommender Systems}

In some of the application scenarios discussed in the next sections, \seq recommender systems represent a special form of context-aware recommender systems.
In session-based recommendation, the users' short-term intents, which can be estimated from their very last actions, can represent an important piece of context information to be taken into account when recommending \cite{jannach15adaptation}.

Time-aware recommender systems (TARS) usually consider time information that is associated with past user actions to adapt the recommendations accordingly, see \cite{campos14time} for an overview. TARS share a number of commonalities with \seq recommenders, e.g., in terms of how we can compare different approaches in offline settings. The focus of \seq recommenders is however often less on the exact point of time of the past user interactions, but on the sequential order of the events. Furthermore, a number of proposals on time-aware recommenders mainly rely on the matrix completion problem setting when modeling temporal dynamics \cite{koren09cf}.



\paragraph{Research in Other Related Fields}
Some aspects of sequence-aware recommender systems were finally explored also in neighboring fields. Examples are the problem of \emph{query suggestion} in the field of information retrieval or the problem of \emph{interest drift} in the more general field of user modeling.
In this paper, we concentrate on works where the recommendation problem itself is the main focus, in contrast to works that, e.g., aim to develop methods to capture changes in the user preferences over time. When searching for papers to consider in our survey, we therefore used a corresponding search string and selection strategy when we queried a digital library, as will be described in more detail in Section \ref{subsec:categorization-existing-works}.
\section{A Categorization of \SeqA Recommendation Tasks}
\label{sec:categorization}
We identified four main goals in the academic literature that can be achieved with the help of \seq recommender systems in different application scenarios:

\begin{enumerate}
\item Context Adaptation
\item Trend Detection
\item Repeated Recommendation
\item Consideration of Order Constraints and Sequential Patterns
\end{enumerate}

\noindent We will discuss these four categories in more detail in the next sections and will then also look at typical application domains for \seq recommenders.
Note that all types of problem settings discussed next are based on the same formal problem characterization described in Equation \ref{eq:optimization-problem}, but require specific algorithmic approaches that use the sequence information in the input datasets (see Section \ref{sec:algorithms}). The problems are also not mutually exclusive and multiple aspects (e.g., trends and repetitions) can be considered in parallel, as was done, for example, in \cite{JannachLudewigLerche2017umuai} for the e-commerce domain.

\subsection{Context Adaptation}
\label{sec:context-adaptation}
In many domains, the relevance of a recommendable item not only depends on the users' general preferences, but also on their current situation and their short-term intents and interests.
Context-aware recommenders take such additional types of information 
into account. Typical contextual factors in the literature include the user's geographical position, the current weather, or the time of the day \citep{adomavicius2011context}. Context factors like these are examples of what is called the \emph{representational context} \citep{dourish04context}, which is defined by a predefined set of ``observable'' context variables.

Contextual factors like the user's current shopping intent in an e-commerce setting or their current mood are however not directly observable. 
These types of information, which represent what is called the user's \emph{interactional context}, therefore have to be derived from the users most recent actions and eventually on behavioral patterns of the user and the community as a whole \cite{hariri12context,natarajan13app}.
Considering interactional context factors is particularly important for systems where there are many new or \emph{anonymous} users. Since no historical data is available about their past preferences, it is important to make full use of interactional context information, as  representational context information can only help to partition anonymous users into coarse-grained categories, without any real personalization \cite{garcin13news}.
Overall, understanding the users' situation and goals and making context-adapted recommendations from past interaction data represents a main goal of \seq{} recommender systems.

\paragraph{Categorization based on importance of long- and short-term interactions}
Depending on the availability of historical data for individual users and the importance of focusing on the most recent interactions,
we can differentiate between the following \emph{types} of context-adaptation situations.

\begin{itemize}
\item \emph{Last-$N$ interactions based recommendation:} In these scenarios, only the last $N$ user actions 
    are considered. 
    A typical problem setting 
    is that of predicting the next location (or \emph{check-in}) in a location-aware recommender system \citep{cheng13where,lian13checkin,liu16next}. The reason to limit oneself to the last actions could be that not many past interactions of that type exist or that the other previous actions of the same type (e.g., check-in events) are not predictive for the next action.
\item \emph{Session-based recommendation:} In this problem scenario only the last sequence of actions of a user is known and this sequence of actions is limited to a \emph{session}, i.e., a limited period of time when the user interacted with the site. 
    Typical application examples include news recommendation \citep{garcin13news}, e-commerce, video and classified advertisement recommendation \cite{hidasi16feature}.
\item \emph{Session-aware recommendation:} Finally, there are situations in which we have knowledge both about the users' actions in the last session \emph{and} about their past behavior. This type of problem setting occurs if we have returning customers that can be identified.
    In this situation, a \seq recommender system can be based on a combination of long-term and short-term interest models, e.g., in e-commerce settings or for app recommendation \citep{baezayates15next,hariri12context,jannach15adaptation,quadrana17personalizing}.
\end{itemize}

Note that our problem definition in Equation \ref{eq:optimization-problem} covers all three scenarios, i.e., the output is a ranked list of items. In the case of a \emph{session-aware} adaptation problem, the underlying sequence dataset $D$ is however usually split into two components, 
where one that contains the older interactions is used for building a long-term model, and the other is used to consider short-term user intents.
How the different models are learned or combined then depends  
on the specific algorithmic approach that is used to maximize the utility function $u$, which might for example return higher scores when recommendations are a mix of familiar and novel items for the user.


\paragraph{On utility functions and recommendation purposes.}
Most of the papers on context-adaptation problems in the literature do not make explicit statements about the characteristics of the utility function 
in the application scenarios they considered. As mentioned above, they in most cases implicitly define the goal through the evaluation procedure and aim to predict hidden elements of a given user session. 
Thereby, they implicitly assume that this next action is in some sense the best recommendation for the given purpose.

The task of recommenders in context-adaptation scenarios can most often be characterized as ``find matching items'' for a given session beginning, without any further explicit specification of what represents a good recommendation.
In some works -- and in practical environments -- a number of more specific purposes can be identified, see also \cite{JannachAdomavicius2016purpose}.
The task of a recommender can be, for example, to create a list of \emph{alternatives} for the currently inspected items (similar items).
In other applications, in contrast, the task can be to determine \emph{complements}, e.g., accessories to a main shopping item in e-commerce.
In yet other application domains the recommendations should represent suitable or logical \emph{continuations} of either the current session (e.g., next-track music recommendations) or the user's longer term behavior (e.g., next-basket recommendations).
Finally, we can differentiate if the user is assumed to pick one of the recommendations (e.g., one alternative in e-commerce scenarios), or consider all of them together (e.g., playlist recommendation for audio and video streaming).
This latter scenario was recently addressed in \cite{Song2017When} for the news domain. In their work, the authors model the user's expected utility of an item during the course of session and try to diversify the recommended content within a session accordingly.

\subsection{Trend Detection}
The detection of trends in a given sequence dataset is another potential, but less explored, goal that can be accomplished by \seq recommenders. We can distinguish between the following types of information that can be extracted from sequential log information to be used in the recommendation process.

\begin{itemize}
\item \emph{Community trends.} Considering the popularity of items within a user community can be important for successful recommendations in practice, e.g., in streaming media recommendation \cite{Gomez-Uribe:2015:NRS:2869770.2843948}. Since the popularity of items can change over time in different domains, \seq recommenders can aim to detect and utilize popularity patterns in the interaction logs to improve the recommendations. Such trends can be long-term (e.g., things becoming outdated or out-of-fashion over time), seasonal, or reflect short-term and one-time popularity peaks. In the fashion domain, for example, considering the community trends of the last few days can represent a successful strategy when selecting items for recommendation \cite{JannachSAC2017ECommerce}.
\item \emph{Individual trends.} Changes in the interest in certain items can also happen at an individual level. These interest changes can be caused when there is a ``natural'' interest drift, e.g., when users grow up, or when their preferences change over time, e.g., due to the influence of other people, due to exceptional events, or when they discover something new.
In the news domain, for example, individual interests change over time and are influenced by global and local news trends \cite{Liu:2010:PNR:1719970.1719976}.
Another example problem is the task of modeling the dynamics of the musical taste of users \cite{moore13taste}.


\end{itemize}

\subsection{Repeated Recommendation}\label{sec:repeated}
In some application domains, recommending items that the user already knows or has purchased in the past can be meaningful. Such scenarios are not considered at all in the traditional matrix completion setup.
We can identify the following categories of repeated recommendation scenarios.
\begin{itemize}
  \item \emph{Identifying repeated user behavior patterns.} Past interaction logs can be used by \seq recommenders to identify patterns of repeated user behavior. A typical application example could be the repeated purchase of consumables, like printer ink.
      Such patterns can be both mined from the behavior of individual users, as in \cite{zhao12intervals,zhao14clusters,wang13opportunity}, or the community as a whole.
      Repeated user actions are particularly relevant for app recommendation problems. In this context, patterns of repeated user behavior can be used to provide shortcuts to applications that are frequently launched in a certain sequence by the user. An example is to suggest to launch the ``e-mail'' or ``calls'' app after opening the ``contacts'' app. The general goal here is to enhance the user experience with the device \cite{baezayates15next,lu14mobile,natarajan13app}.

  \item \emph{Repeated recommendations as reminders.} In a different scenario, repeated recommendations can help to \emph{remind} users of things they found interesting in the past. Depending on the domain, these reminders could relate to objects that the user has potentially forgotten (e.g., an artist that she or he liked in the past), or to objects that the user has recently interacted with \cite{Lerche2016Reminder}. The latter scenario is particularly relevant in e-commerce, and the recommendation of recently-inspected items is common on platforms like Amazon.com.
\end{itemize}

Note that from the viewpoint of our problem characterization in Section \ref{sec:characterizing}, the two scenarios are identical. In the first case, however, there is often an underlying ``logical'' reason why an item should be recommended again, which is not the case for scenarios of the latter class, i.e., the recommendation of assumedly ``forgotten'' or recently relevant items.

In both mentioned scenarios, besides the selection of items to repeatedly recommend, a \seq recommender has to reason about the \emph{timing} of the recommendations. In the reminding scenario in e-commerce, the time frame to remind users of previously seen items might be narrow and objects may become obsolete soon, e.g., if they were not purchased after a few view events. Nonetheless, always reminding users of items they have inspected in the last session might be inappropriate if the user's current shopping intent does not match that of the previous session.
In the context of the recommendation of consumables, items can be repeatedly recommended after longer periods of time, e.g., weeks or even months. Proper timing can however still be important and such types of repeated recommendations share similar problems as \emph{proactive} recommenders \cite{DaliBetzalel:2015:PMT:2792838.2799672}, e.g., that their recommendations might interrupt the user at the wrong time.

\subsection{Consideration of Order Constraints and Observed Sequential Patterns}
In Section \ref{sec:context-adaptation} on short-term context adaptation, we have discussed different tasks of \seq recommender systems in which the order of objects -- either in the logs, in the current session, or in the recommendations -- can play a role. Considering such orderings can be central to \seq recommendation tasks also outside the context of the user's last session, which is why we discuss this aspect in some more depth in this section.

Specifically, there are two types of information about sequentiality one can additionally consider when determining a suitable order of the recommendations.

\begin{enumerate}[label=(\alph*)]
\item First, there can be ``external'' domain knowledge in the form of strict or weak ordering constraints that prescribes the ordering.
In the domain of recommending a sequence of learning courses, for example, there might be strict requirements regarding the order of different courses that have to be considered by the recommender, e.g., when one cannot attend one course before another one was completed \cite{xu16personalized,Parameswaran:2011:RSC:2037661.2037665}. In the domain of movie recommendation, in contrast, it might be reasonable to recommend a sequel to a movie only after a user has watched the preceding episode. Such a constraint is however not necessarily strict.
\item Second, one can try to identify such sequential consumption patterns from the user behavior, and, e.g., automatically infer that users who watched a certain movie later on watched its sequel.
From a technical perspective, sequential pattern mining techniques have, for example, been applied in different application domains of recommenders, e.g., for predicting the next navigation actions of users on websites \citep{mobasher02sequential,nakagawa03impact} or the find next tracks to play in music recommendation problems \citep{bonnin14playlists}.
\end{enumerate}

When considering these factors that can influence the ordering, a number of application-specific variations might have to be considered, among them the following.
\begin{itemize}
\item \emph{Importance of the order of the recommendations.} Depending on the specifics of the application scenario and the goals of the recommendation service, the order of the recommended items can be relevant or not, even in the same domain. In the context of next-track music recommendation, one can try to determine a playlist continuation where all elements are generally a good fit for the current listening session \cite{jannach15playlist}. In contrast, one could however also try to make sure that also the transitions between the tracks are smooth or that the resulting playlist has certain characteristics, e.g., a continuous increase of the tempo \cite{maillet09playlist}.

\item \emph{Importance of the exact order of the past events.} Similarly, the exact order of the past events may or may not be relevant for the recommendation task. Again in the domain of the music recommendation, one can try to look for sequential patterns in past sessions as in \cite{hariri12context} or simply consider track co-occurrence patterns, as done in the neighborhood-based approach in \cite{bonnin14playlists}. In cases where the order is relevant, one might additionally consider the age of the individual events in the log and reduce the importance of older events.

\item \emph{Existence of implicit order constraints.} When recommending complements (e.g., accessories for a given item), there might be implicit constraints of what can be reasonably recommended and these constraints can depend on the specific domain or product category. One can recommend a memory card as an accessory to a camera, 
    but not the other way round. For other categories, however, recommendations in both directions can be plausible.

\end{itemize}

In general, besides the identification of sequential patterns from past data to select and rank items, an additional task for the recommender is to ensure that such strict and weak order constraints are respected in the resulting lists.

\subsection{Categorization of Existing Works}
\label{subsec:categorization-existing-works}
In this section we will classify existing research works according to our categorization scheme in order to obtain a better understanding of which aspects of \seq{} recommenders are comparably well explored and which areas need further investigation.

We selected the papers to be considered in this categorization as follows. We issued a query to the ACM Digital Library using a set of relevant search terms\footnote{The exact query was ``sequence sequential trend next basket bundles repeat session order + recommend''}, sorted the first search 1,000 results by relevance and manually inspected the abstracts of the first 100 papers. If a paper made explicit use of sequences of user actions it was considered relevant and its referenced papers were scanned as further potential candidates to consider. Table \ref{tab:categorization-goals-tasks-1} shows a summary of the papers that were finally considered relevant for this work.

In the table, the first column shows the context-adaptation type of each approach according to our categorization. 
The second column indicates which form of ordering constraints were considered; the third column mentions the main application domain addressed in the papers.

\begin{table}[t]
\scriptsize
\caption{Categorization of works regarding Context-Adaptation Type, Order Constraints and Domain.}
\label{tab:categorization-goals-tasks-1}
\begin{tabular}[t]{@{}llll@{}}
\toprule
\textbf{Paper} & \textbf{Context} & \textbf{Ord.} & \textbf{Domain}\\\midrule
\citet{baezayates15next} & SA & I & APP\\
\citet{chen12embedding} & LI & WI & MUS\\
\citet{chen13multispace} & LI & WI & MUS\\
\citet{cheng13where} & LI & I & POI\\
\citet{chou16cold} & LI & I & MUS\\
\citet{feng15personalized} & LI & I & POI\\
\citet{garcin13news} & SB & I & NEWS\\
\citet{grbovic15prod2vec} & LI & I & EC\\
\citet{greenstein-messica17session} & SB & I & EC\\
\citet{hariri12context} & SB & WI & MUS\\
\citet{he09query} & SB & I & QRY\\
\citet{he16fusing} & LI & I & EC,POI\\
\citet{he16inferring} & LI & I & POI\\
\citet{hidasi16feature} & SB & I & EC,VID\\
\citet{hidasi_session-based_2016} & SB & I & EC\\
\citet{hosseinzadeh15adapting} & LI & I & MUS\\
\citet{hsueh08mining} & LI & I & EC\\
\citet{jannach15adaptation} & SA & I & EC\\
\citet{jannach15playlist} & SA & WI & MUS\\
\citet{JannachSAC2017ECommerce} & SA & --- & EC\\
\citet{Lerche2016Reminder} & SA & I & EC\\
\citet{letham13sequential} & LI & I & EC\\
\citet{lian13checkin} & LI & I & POI\\
\citet{lim15personalized} & LI & S & POI\\
\citet{liu16next} & LI & SI & POI\\
\citet{liu16unified} & LI & I & POI\\
\citet{lu14mobile} & SB & I & APP\\
\citet{maillet09playlist} & --- & W & MUS\\
\citet{mcfee11nlp} & LI & WI & MUS\\
\citet{mobasher02sequential} & SB & I & WWW\\
\citet{moling12optimal} & SB & --- & MUS\\
\citet{moore13taste} & LI & WI & MUS\\
\citet{nakagawa03impact} & SB & I & WWW\\
\citet{natarajan13app} & SA & I & MUS,APP\\
\end{tabular}
\begin{tabular}[t]{@{}llll@{}}
\toprule
\textbf{Paper} & \textbf{Context} & \textbf{Ord.} & \textbf{Domain}\\\midrule
\citet{Parameswaran:2011:RSC:2037661.2037665} & --- & S & CS \\
\citet{pauws06fast} & --- & S & MUS\\
\citet{quadrana17personalizing} & SA & I & EC,VID\\
\citet{reddy16learning} & --- & W & CS\\
\citet{rendle10FPMC} & LI & I & EC\\
\citet{rudin11sequential} & LI & I & EC\\
\citet{shani05mdp} & LI & I & EC,WWW\\
\citet{soh17deep} & SB & I & OTH\\
\citet{song15next} & SB & I & EC,VID\\
\citet{song16multirate} & SB & I & NEWS\\
\citet{sordoni15hierarchical} & SB & I & QRY\\
\citet{tagami15modeling} & LI & I & ADS\\
\citet{tavakol14fmdp} & SB & I & EC\\
\citet{trevisiol14cold} & SB & I & WWW,NEWS\\
\citet{turrin15large} & SB & I & MUS\\
\citet{twardowski16contextual} & SB & I & EC\\
\citet{vasile16meta} & SB & WI & MUS\\
\citet{wang13opportunity} & LI & I & EC\\
\citet{wang15basket} & LI & I & EC\\
\citet{wu13karaoke} & SB & WI & MUS\\
\citet{xiang10temporal} & SA & I & OTH\\
\citet{xu16personalized} & --- & S & CS\\
\citet{yan11context} & SB & I & QRY\\
\citet{yap12sequential} & LI & I & WWW\\
\citet{yu12sequential} & LI & S & OTH\\
\citet{yu16dynamic} & LI & I & EC\\
\citet{zang10non-redundant} & LI & I & EC\\
\citet{zhang13sequential} & LI & I & ADS\\
\citet{zhang14lore} & LI & I & POI\\
\citet{zhao12intervals}   & LI & I & EC\\
\citet{zhao14clusters}    & LI & I & EC\\
\citet{zheleva2010www}    & SA & ---  & MUS\\
\citet{zhou04intelligent} & LI & I & WWW\\
\citet{zidmars01temporal} & LI & I & WWW\\
\end{tabular}

\begin{tabular}{llllllll}
\midrule
\multicolumn{8}{@{}p{\linewidth}@{}} {\scriptsize \textbf{Context} SA: Session-aware, SB: Session-based, LI: Last-$N$ interactions; \textbf{Order Constraints} W: Weak, I: Inferred, S: Strong; \textbf{Application domain} APP: App recommendation, MUS: Music domain, QRY: Query recommendation, EC: E-Commerce domain, CS: Learning courses recommendations,
AD: Advertisements, WWW: Web navigation recommendation, OTH: Others}\\
\bottomrule
\end{tabular}
\end{table}

\subsubsection{Categorization based on Context-Adaptation Type} 
\begin{figure*}[h!t]
    \centering
    \includegraphics[width=0.6\textwidth,clip]{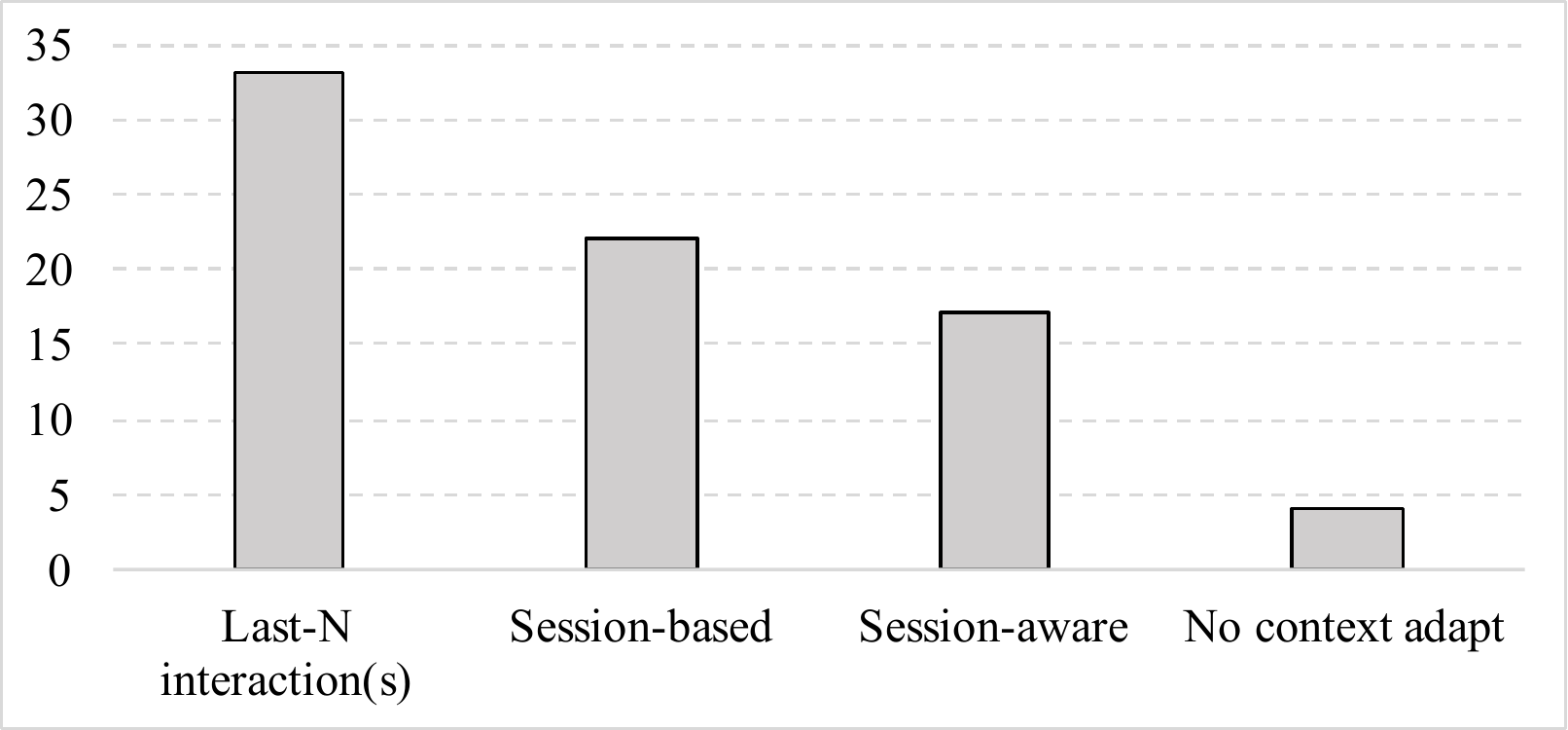}
    \caption{Distribution of types of context adaptation in the analyzed research works.}
    \label{fig:histogram-context-adaptation-types}
\end{figure*}

We differentiate between session-aware, session-based, and last-interaction adaptation approaches.
Figure \ref{fig:histogram-context-adaptation-types} shows the number of research works that fall into the different categories.

We can see that a large fraction of the papers focuses only on the very last visited object when determining the next object to be recommended. A slightly smaller number of papers consider session-based recommendation scenarios. 
However, we can observe increased interest in such session-based recommendation problems in the recent years, which is fueled both by the industry\footnote{For example, a major e-commerce platform provided a new dataset in the context of the 2015 ACM RecSys Challenge about session-based recommendation.} and by the emergence of new deep learning based sequence-learning techniques, as we will discuss in Section \ref{sec:algorithms}. In fact, more than half of the papers on session-based and session-aware recommendation were published starting from 2014.
Finally, even fewer papers also take the long-term preferences of the user into account. While short-term user intents might in many cases be more relevant than long-term preferences, some works -- like the ones by \cite{jannach15adaptation,quadrana17personalizing} -- show that considering long-term preferences can be important to achieve more accurate recommendations.

The limited number of works on session-aware systems might partially be due to the lack of publicly available benchmark datasets. Overall, however, the results shown in Figure \ref{fig:histogram-context-adaptation-types} indicate more research is required to better understand the interplay between long-term and short-term preferences in \seq recommenders.

\subsubsection{Order Constraints} 
As the column labelled ``Ord.'' in Table \ref{tab:categorization-goals-tasks-1} shows, most existing works in sequence-aware recommendation rely on implicitly derived (and correspondingly weak) ordering constraints and only a few works address scenarios where explicit and strong order constraints have to be considered. This can again be caused by the focus of the research community on certain application domains like media (movie) recommendation \cite{jannach2012recommender}, where ordering constraints exist -- consider movie sequels or follow-up news stories -- but are often not considered in the corresponding algorithmic approaches.

\subsubsection{Categorization based on Domain}
\begin{figure*}[h!t]
    \centering
    \includegraphics[width=0.6\textwidth,clip]{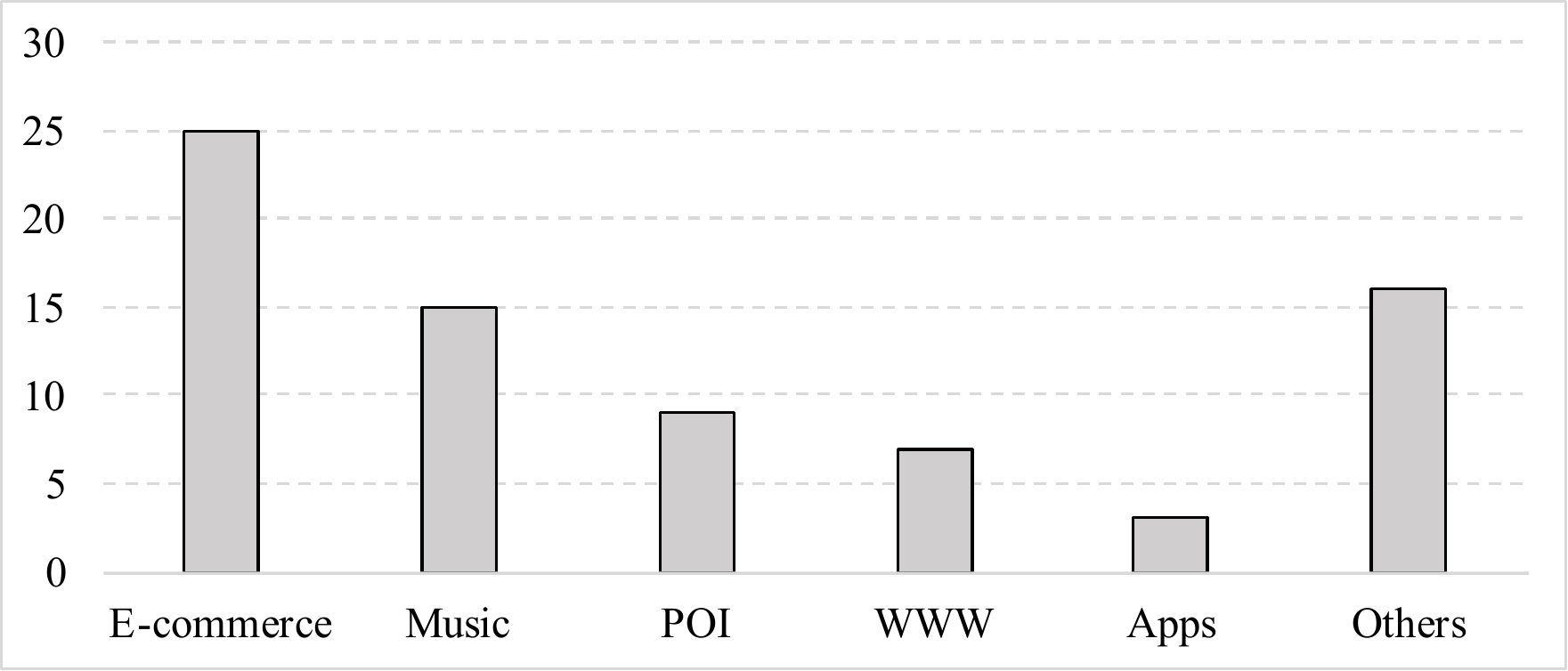}
    \caption{Distribution of research works per application domain.}
    \label{fig:histogram-per-domain}
\end{figure*}

Figure \ref{fig:histogram-per-domain} finally shows in which application domain \seq{} recommenders were investigated. 
In the following, we give examples of typical research works in each domain.

\emph{E-commerce} is the most often investigated domain, which is not surprising as session-based recommendation and session-aware recommendation, as defined above, are common tasks in e-commerce scenarios and investigated, e.g., in  \cite{hidasi_session-based_2016,hidasi16feature} or \cite{jannach15adaptation}.
Other problem settings in the e-commerce domain include next-basket recommendation \cite{rendle10FPMC,wang15basket,yu16dynamic} and the problem of \emph{reminding} users, e.g., in \citep{zhao12intervals,zhao14clusters} and \citep{Lerche2016Reminder}.

\emph{Music recommendation} is another ``natural'' application domain of \seq recommenders. The consumption of music is often session-based and the listener's interest can change strongly from one session to another. The user experience can furthermore be influenced by the order in which the tracks are played, i.e., weak ordering constraints can exist between the tracks. Such constraints can either be explicitly given by the user \citep{pauws06fast} or can be inferred from listening logs of a user community as done, e.g., in \citep{chen12embedding} and \citep{jannach15playlist}.
Constraints can be inferred on the basis of the transitions between pairs of songs \cite{chen12embedding,chen13multispace}, or based on metadata and other attributes of tracks that are usually ``played together'' \cite{vasile16meta,jannach15playlist}. This knowledge can then be leveraged by \seq recommenders to generate playlists or plausible continuations of listening sessions, see, e.g., \cite{jannach15playlist} or \cite{turrin15large}.

\emph{POI recommendations} make predictions on the user's next location or make recommendations for the next place to visit. Moving between places is a sequential process, where the user's movements are usually limited by distance, time, or budget constraints. \Seq recommenders have been applied in this context in different ways to predict the next user location, e.g., based on the user's current location \cite{cheng13where,he16fusing}. Considering several past user locations has shown to be helpful, for example, for travel planning \cite{lim15personalized}, or when predicting which place the user will most probably visit at a specific time \cite{liu16unified}.

\emph{Web navigation prediction} is an early application area of \seq recommenders
\citep{mobasher02sequential,nakagawa03impact,zhou04intelligent}. Web browsing is usually a sequential process and next-page visit predictions can help to recommend users interesting links that fit their current browsing session or to pre-load web pages.

\emph{App recommendation} or app usage prediction is a more recent application field of \seq recommenders, where considering the user's current context is crucial. A typical example of a research work is described in \citep{baezayates15next}, where repeated usage patterns are mined from activity logs to pre-fetch applications or to make contextual suggestions on which app to use.

\emph{Others.} Finally, there are a number of other application domains of \seq recommenders which have however only be explored by a few research works. Examples include \textit{advertisement} recommendation \citep{zhang13sequential}, news recommendation \citep{garcin13news}, job posting recommendation \cite{quadrana17personalizing}, the recommendation of videos in streaming platforms \citep{hidasi16feature,quadrana17personalizing}, query recommendation \citep{baezayates04query,he09query,sordoni15hierarchical,cao08context}, or the recommendation of next activities to add during workflow modeling \citep{JannachJugovacEtAl2016}.

\subsubsection{Specific Tasks of \Seq{} Recommendation}

Table \ref{tab:categorization-goals-tasks-2} lists works that considered the problems of repeated recommendations and trend detection.


\begin{table}[h!t]
\caption{Categorization of works regarding Repeated Recommendation and Trend Detection.}
\label{tab:categorization-goals-tasks-2}
\footnotesize
\begin{tabular}{l l}
\toprule
\textbf{Repeated Recommendation} & \\
\tabitem Find items for repeated recommendation & \cite{JannachSAC2017ECommerce,Lerche2016Reminder} \\
\tabitem Find timing for repeated recommendation & \cite{liu16next,liu16unified,wang13opportunity,zhao12intervals,zhao14clusters} \\\midrule
\textbf{Trend detection} & \\
\tabitem Detect individual trends & \cite{moore13taste,wang13opportunity} \\
\tabitem Detect community trend & \cite{wang13opportunity,song16multirate} \\
\tabitem Detect seasonal trends & \cite{wang13opportunity} \\
\bottomrule
\end{tabular}
\end{table}

Repeated recommendations, while useful in practical applications \cite{Lerche2016Reminder}, were rarely the focus of researchers in the context of \seq{} recommenders.
Similarly, the detection of trends has also not been investigated to a large extent, even though a recent work suggest that taking in particular short-term trends can be important in the e-commerce domain \cite{JL2017SAC}. A few works consider community trends and some investigate seasonal aspects. Interestingly, only two works analyzed here considered individual consumer trends and potentially outdated user tastes \cite{moore13taste,wang13opportunity}.

Finally, there are only few works that explicitly mention that the goal is to recommend \emph{similar} items \cite{jannach15playlist,moore13taste,greenstein-messica17session} or \emph{complements} \cite{tavakol14fmdp}.
A large number of papers in the literature focus explicitly on \textit{next-item} recommendation \cite{baezayates15next,he09query,hidasi16feature,lu14mobile,mobasher02sequential,nakagawa03impact,natarajan13app,song15next,sordoni15hierarchical,zhang13sequential,zhou04intelligent} and \textit{list continuation} \cite{hariri12context,jannach15playlist,maillet09playlist,yu12sequential}.
The problem of \textit{next-basket} recommendation in e-commerce was also the focus of only a few works \citep{rendle10FPMC,wang15basket,yu16dynamic,lian13checkin}.
\section{Algorithms for \SeqA Recommender Systems}
\label{sec:algorithms}
\newcommand\applicationexamples{Application examples}

We can identify three main classes of algorithms in the literature that are used for the extraction of
patterns from the sequential log of user actions: \textit{sequence learning}, \textit{\seq matrix-factorization} and \textit{hybrids}. Each of these classes can be further decomposed into subcategories, leading to the \emph{taxonomy of algorithms} in \autoref{tab:algo_taxonomy}. A few comparably uncommon technical approaches are listed under the category ``Others''.

The majority of the reviewed works rely on some form of sequence learning methods. 
Such approaches are a natural choice for the given problem and frequent pattern mining techniques, for example, have been applied for a long time, e.g., for the prediction of user navigation patterns on websites \cite{mobasher02sequential,nakagawa03impact}.
Other approaches, in particular the ones based on recurrent neural networks and distributed item representations, were only recently successfully transferred from domains like Natural Language Processing to sequential recommendation problems \cite{hidasi16feature,grbovic15prod2vec}.
We discuss the different approaches in more detail in the next sections.


\begin{table}[t]
\footnotesize
\centering
\setlength{\tabcolsep}{3pt}
\caption{Taxonomy of algorithms for \seq recommender systems.}
\label{tab:algo_taxonomy}
\begin{tabular}{p{2.5cm}P{2.5cm}P{2.5cm}p{5cm}}
\toprule
\textbf{Class} & \textbf{Sub-class} & \textbf{Family} & \textbf{Examples of research works} \\ \midrule
Sequence Learning & \multirow{2}{2.5cm}{Frequent Pattern Mining} & Frequent itemsets &
\cite{jannach15adaptation,mobasher02sequential,nakagawa03impact,turrin15large,zang10non-redundant,hsueh08mining}\\\cmidrule(r){3-4}
&& Frequent sequences & \cite{lu14mobile,mobasher02sequential,nakagawa03impact,rudin11sequential,yap12sequential,zhou04intelligent}\\\cmidrule(r){2-4}
& Sequence modeling & Markov Models & \cite{he09query,garcin13news,hosseinzadeh15adapting,mcfee11nlp} \\ \cmidrule(r){3-4}
& & Reinforcement Learning & \cite{moling12optimal,shani05mdp,tavakol14fmdp} \\ \cmidrule(r){3-4}
& & Recurrent Neural Networks & \cite{zhang13sequential,hidasi_session-based_2016,hidasi16feature,liu16unified,sordoni15hierarchical,song16multirate,twardowski16contextual,yu16dynamic,du16recurrent,soh17deep}\\ \cmidrule(r){2-4}
& Distributed item representations & Latent Markov embeddings & \cite{chen12embedding,chen13multispace,wu13karaoke,feng15personalized}\\ \cmidrule(r){3-4}
& & Distributional embeddings &
\cite{djuric14hidden,grbovic15prod2vec,tagami15modeling,vasile16meta,baezayates15next,reddy16learning,zheleva2010www}\\ \cmidrule(r){2-4}
& Supervised models w/ sliding window & & \cite{zidmars01temporal,baezayates15next,wang15basket}\\ \midrule
Matrix factorization & & & \cite{twardowski16contextual,yu12sequential,zhao12intervals,zhao14clusters}\\ \midrule
Hybrid methods & Factorized Markov Chains & & \cite{rendle10FPMC,cheng13where,he16inferring,he16fusing,lian13checkin}\\ \cmidrule(r){2-4}
& LDA/Clustering w/ sequence learning & & \cite{hariri12context,song15next,natarajan13app}\\ \midrule
Others & Graph-based & & \cite{xiang10temporal,liu16unified,trevisiol14cold,zhang14lore}\\ \cmidrule(r){2-4}
& Discrete optimization & & \cite{jannach15playlist,lim15personalized,pauws06fast,xu16personalized}\\
\bottomrule
\end{tabular}
\end{table}

\subsection{Sequence Learning}
\label{sec:sequence_learning}
Sequence learning methods are useful in application domains where the data to be analyzed has an inherent sequential nature, like in Natural Language Processing, time-series prediction, DNA modeling, and, as the focus of this work, \seq recommendation.


\subsubsection{Frequent Pattern Mining (FPM)}
\label{sec:frequent_patterns}

\paragraph{Methods}
Frequent Pattern Mining (FPM) techniques were originally developed to discover user consumption patterns within large transaction databases. Early Association Rule Mining approaches \citep{agrawal94association} focused on identifying items that frequently co-occur in the same transaction, regardless of the order of their appearance. Later on, Sequential Pattern Mining \cite{agrawal95mining} techniques were developed that considered item co-occurrences only as a pattern when the items appeared in the same order. An extreme case finally are Contiguous Sequential Patterns, which require that the co-occurring items are adjacent in the sequence of actions within a transaction.

Technically, in all approaches the patterns are mined in an offline process and usually translated into a set of association rules or another compact form of knowledge representation, see, e.g., \cite{mobasher02sequential,yap12sequential}. Usually, the resulting rules have values attached (e.g., \emph{confidence} and \emph{support}) that express their strength. In the prediction phase, we are given a partial transaction (e.g., an item that a user has recently bought) for which we seek additional items. These items are then determined by scanning the database for matching rules and by applying them on our partial transaction. In recommendation scenarios, the most simple approach is to determine only pairwise item co-occurrence frequencies in order to implement buying suggestions of the form ``Customers who bought~\dots~also bought''.

\paragraph{\applicationexamples}
In one of the earlier works on \seq recommenders \cite{mobasher02sequential,nakagawa03impact}, researchers compared Association Rules (AR), Sequential Patterns (SP) and Contiguous Sequential Patterns (CSP) for a web usage mining scenario, where the problem is to predict a user's next navigation action for page prefetching or for \textit{context-adaptation} in \textit{session-based} scenarios. Technically, they used a fixed-sized sliding window of the current user session in the prediction phase. Given the last $N$ user actions in this window they look up and apply rules of size $N+1$ and rank the recommendations (i.e., the elements of the rule consequents) based on the confidence value of the firing rules.\footnote{A technically similar approach was later on proposed by \citet{zhou04intelligent}.} The obtained results showed that the less constrained patterns (AR and SP) lead to better recommendations, whereas the usage of CSPs was more helpful for the page prefetching task.
Other kinds of sequential patterns, such as closed patterns and negative patterns, were explored too. See, for example, the works of \citet{zang10non-redundant} and \citet{hsueh08mining} for \textit{context-adaptation} based solely on the last few user actions. 

When using frequent pattern approaches, personalization is obtained by matching the activity of the user with the pre-extracted patterns. In a more recent work, \citet{yap12sequential} propose to further personalize the method and to weight the patterns according to their estimated relevance for the individual user. They designed three schemes to determine personalized pattern relevance scores and compared it to a popularity-based method. Their empirical evaluation indicates that applying personalized rule scoring schemes yields more accurate personalized \textit{next-item recommendations} for the target users.

As an example of a comparably recent application domain, \citet{lu14mobile} present the MASP (\emph{Mobile Application Sequential Patterns}) mining method for the problem of providing \textit{context-adaptation} to smartphones by predicting the user's next used app. In their approach, transactions are composed of sequences of app usages which are annotated with the location of the user at each time. The MASP-mine algorithm takes into account both the user movements and app launches to discover the sequential patterns. At prediction time, the single pattern with the maximal support value that matches the recent user movements and activity is used to predict the app that is launched next.

\paragraph{Discussion}
Frequent pattern mining techniques are well-explored and also easy to implement and interpret.
The main drawbacks include the limited scalability of some of the approaches and, probably more importantly, the problem of finding suitable threshold values for the offline mining task. Usually, the main parameter is the minimum support value. If it is set too low, too many (often noisy) patterns are identified; if it is set too high, only rules for the most frequently occurring items will be found. To deal with the problem, one can start to search for the highest-quality rules 
in a database that was created with a fixed minimum-support threshold, and iteratively relax the quality constraints until a matching rule is found. 
Alternatively, different algorithm variants were proposed 
that use multiple minimum support values or ``adjusted'' confidence scores \cite{rudin11sequential}.

Another common challenge when using frequent pattern mining techniques is to decide between the different variants (AR, SP, CSP). Determining sequential patterns (SP, CSP) is often not only computationally more expensive given the more strict type of rules to be mined\footnote{See \cite{agrawal94association,agrawal95mining} for a detailed analysis on the computational complexity of FPM methods.}, but it can also easily lead to much smaller rule bases.
This in turn can result in situations at prediction time where no matching rule is found. Which method works best can furthermore depend on the application domain. The ordering of the events can be very important, e.g., in the context of query or app recommendation \cite{zhou04intelligent,lu14mobile}. In other problem settings, e.g., web page recommendation or next-track music recommendation, considering the item orderings might result in too small rule bases or have only small positive effects \cite{BonninJannach2013} that probably do not outweigh the additional computational complexity. Finally, in some domains, simple co-occurrence patterns were despite their simplicity employed with good success, e.g., in e-commerce and music recommendation \cite{jannach15adaptation,turrin15large,JannachLudewig2017RecSys}.

\subsubsection{Sequence modeling}
\label{sec:seq_models}

\paragraph{Methods}
The input to \seq recommenders, i.e., the ordered and often time-stamped log of past user actions, can be considered as a \emph{time series} with discrete observations. Therefore, in many cases existing and sometimes complex time series prediction methods could in principle be applied.
In RS applications, however, the time-stamps are often merely used to sort the actions\footnote{The work of \citet{du16recurrent} represents an exception in that respect.}, which allows us to apply ``simpler'' sequence models, that do not necessarily consider the complex underlying temporal dynamics of the observed sequences of actions.

In general, sequence modeling techniques aim to learn models from past observations to predict future ones, which are in our case user actions.
Sequence modeling methods for \seq recommendation mainly belong to three categories: Markov Models, Reinforcement Learning and Recurrent Neural Networks (see \autoref{tab:algo_taxonomy}).

\begin{itemize}
\item \emph{Markov Models} consider sequential data as a stochastic process over discrete random variables (or states). The Markov property limits the dependencies of the process to a finite history. For example, in first-order Markov Chains (MCs) the transition probability of every state depends only on the previous state. Higher-order MCs use longer temporal dependencies to model more complex relationships between the states\footnote{See \cite{fink2014markov} for details on Markov Models for pattern recognition.}. In \seq recommender systems, the Markov property translates into assuming that the next user actions depend only on a limited number of the most recent preceding actions.


\item \emph{Reinforcement Learning} (RL) techniques learn by interacting with the environment, and are sequential in nature. In a recommendation scenario, the interaction consists of a recommendation of an item to the user (the \emph{action}) for which the system then receives a feedback (the \emph{reward})\footnote{See \cite{kaelbling1996reinforcement} for a comprehensive review on Reinforcement Learning.}.
For instance, in the music domain, the system recommends a song and monitors if the user listens to or skips the recommended song.
In this example, we assign a positive reward if the user listens to the song and zero otherwise.
The problem is typically formulated as a Markov Decision Process (MDP) and the goal of the system is to maximize the cumulative reward computed over a number of interactions.


\item \emph{Recurrent Neural Networks (RNNs)}
are distributed real-valued hidden state models with non-linear dynamics\footnote{See \cite{lipton15critical} for an overview of Recurrent Neural Networks.}. At each time step, the hidden state of the RNN is computed from the current input in the sequence and the hidden state from the last step.
The hidden state is then used to predict the probability of the next items in the sequence. The recurrent feedback mechanism memorizes the influence of each past data sample in the hidden state of the RNN, hence overcoming the fundamental limitation of MCs. RNNs are therefore well suited for modeling the complex dynamics in user action sequences. Variants of RNNs such as LSTM \cite{graves13lstm} and GRU \cite{chung14gru}, by means of their sophisticated hidden dynamics, can model much longer and complex temporal dependencies than other approaches like Hidden Markov Models \citep{ghahramani01hmm}.

\end{itemize}

\paragraph{\applicationexamples{} (Markov Models)}
Markov Chains in most cases cannot be naively applied to \seq recommendation since data sparsity quickly leads to poor estimates of the transition matrices. \citet{shani05mdp} therefore enhance their MC-based approach with several heuristics -- namely skipping, clustering and finite mixture modeling -- to mitigate the impacts of data sparsity.
In their application, the input data consists of purchase records of an online book store and their experiments show the superiority of sequential models over non-sequential ones in predicting what books to recommend \textit{next} given the sequence of the last few user interactions. 
In a different application domain, \citet{mcfee11nlp} use Markov Chain \emph{mixtures} for the problem of music playlist generation, a typical \textit{list continuation} problem, where a mixture is a weighted ensemble of several MCs (uniform, weighted and k-nearest neighbors) whose weights are learned via maximum-likelihood optimization on a training dataset of playlists.

Another challenge when applying MCs usually lies in the choice of the order of the model. In the context of a \textit{session-based} next-query recommendation problem, \citet{he09query} therefore use a mixture of Variable-order Markov Models (VMMs, sometimes called \emph{context trees}\footnote{See \cite{begleiter04vmm} for a comprehensive analysis of algorithms for learning VMMs.}), which use a context-dependent order to capture both large and small Markov dependencies. 
A different approach is adopted by \citet{garcin13news} in the context of a news recommendation application. They assign one predictor (expert) to each context (node) in the context-tree, where each node is associated with a different order of the Markov model.
Each predictor is then trained to predict the article to suggest \textit{next} given the last sequence of user actions.
As the sequence of actions of the user grows, deeper nodes in the tree become active and contribute to the final recommendations.

Hidden-state models, in particular Hidden Markov Models (HMM), address some of the limitations of MC models and are, for example, applied in \cite{hosseinzadeh15adapting} for the \textit{contextual next-track} music recommendation problem.
In HMMs, each hidden (or latent) state is a discrete variable associated with a probability distribution over the observed variables. In conformity with the Markov property, every hidden state 
is conditionally dependent only on the previous one \cite{ghahramani01hmm}. There already exist a few applications of HMMs to time-aware collaborative filtering \cite{Sahoo:2012:HMM:2481674.2481689,zhang17modeling}.
In \cite{hosseinzadeh15adapting}, the hidden states is used to model the (unobserved) context of the user. Discrete-valued hidden states are however limited in terms of the contextual information they can store, which to some extent limits their applicability for \seq recommendation.

\paragraph{\applicationexamples{} (Reinforcement Learning)}
Reinforcement learning based on MDPs are used in \cite{shani05mdp} and \cite{moling12optimal} for \seq recommendation in on-line shops. These approaches make it possible to tailor recommendations not only to the recent user activity but also to the expected reward (income) for the shop.
Since the state space of MDPs can quickly become unmanageable in realistic scenarios, \citet{tavakol14fmdp} factorize the space over a set of mutually independent item attributes. 
In their application case of a clothing marketplace, dress characteristics such as category, color, or price can be considered; the sequential relationships between attributes are then independently modeled by an MDP to predict the characteristics of the products the user will likely search or buy next.

\paragraph{\applicationexamples{} (RNNs)}

\citet{zhang13sequential} use RNNs for click prediction for online advertisements. At each time step, they train the RNN to predict the next click of the users given their last click and the previous state of the network using a classification loss measure (cross-entropy). 

For the domains of e-commerce and media recommendation, \citet{hidasi_session-based_2016} explore the use of Gated Recurrent Units (GRUs) \citep{chung14gru} -- a variant of RNNs -- for modeling user activity in a session-based scenario. Technically, as in \cite{zhang13sequential}, the model is trained to predict the next item in a sequence given the current one, but different loss functions were used, which at the end led to better performance. 
Later on, \citet{hidasi16feature} included item features 
into the sequence model. The proposed parallel-RNNs model is able to capture the interdependencies between item identifiers and their features, leading to a further improved recommendation accuracy.
\citet{quadrana17personalizing} employ \emph{hierarchical} RNNs for modeling user preferences across sessions. The first level of the hierarchy, which represents the state of the user across sessions, is used to initialize and propagate information to the second level of the hierarchy, which is used to generate recommendations within a session. Transferring the information from prior sessions ultimately led to better recommendations.
\citet{yu16dynamic} finally employ RNNs for \textit{next-basket} recommendation in e-commerce. In their case, the RNN is fed with real-valued representations of the baskets as an input -- i.e., the sets of items the user has interacted with at each step of the sequence -- and trained to rank the items in the next basket.

A number of further technical variants of RNNs and problem encodings were recently proposed for different application domains like e-commerce, online ads, and query or POI recommendation. \citet{twardowski16contextual}, for example, addresses the problem of providing personalized recommendations to \emph{anonymous users} by using RNNs to generate a compact representation of the user's interactional context. Such representation is later combined with item representations to genrete \textit{contextual} recommendations.
\citet{liu16unified} aim at detecting \emph{interest trends} of individual users over time in an RNN-based POI recommender system.
Both \citet{sordoni15hierarchical} and \citet{soh17deep} address the \emph{context adaptation} problem.
\citet{sordoni15hierarchical} propose a generative model for session-based query recommendation based on Hierarchical Recurrent Encoder-Decoder neural networks. The higher level of the hiercarchy encodes the latent intent of the user in the current search session, while the lower level generates next-query suggestions tailored to this intent.
\citet{soh17deep} employ RNNs to represent sequences of user interactions to personalize user interfaces.

\citet{song16multirate} propose a \emph{context adaptation} approach to build a session-aware recommender that can handle long-term (e.g. seasonal) and short-term changes in user preferences in the news domain.
In their work, the authors propose the Temporal Deep Semantic Structured Model to combine user and item features with user temporal features into a joint model. Static features are modeled by several feed-forward neural networks, whereas the temporal features are modeled by a set of RNNs. 

\citet{du16recurrent} address the \emph{list continuation} problem in a wide range of application domains, from taxi drop-off to financial transactions.
Given data with timestamps for the actions, their model combines RNNs with Marked Temporal Point Processes, a mathematical tool that models the distance between subsequent events as a random process.
The idea of modeling the distance between subsequent events as a random process is used also by \citet{wang13opportunity}, who employ a hierarchical Bayesian framework based on hazard models to represent the probability of purchasing a product at a given time, while modeling time-dependent patterns between follow-up purchases at the same time.
Both models allow to compute the utility of recommendations as a function of the sequence of past user actions and of the recommendation timing.


\paragraph{Discussion}
Sequence modeling approaches have been successfully applied for a variety of \seq{} recommendation problems.
Our review however shows that Markov Models in most cases cannot be directly applied in a naive manner for \seq recommendation problems, due to problems of data sparsity and computational complexity, which is why researchers often rely on specific model variants or embed heuristics into the learning process.
Still, it is not always fully clear if such model variants scale to real-world problems.

Deep learning based techniques, on the other hand, have been increasingly explored in the past few years. 
Among the factors that contributed to the recent revival of neural networks are the availability of large data sets for training and the increased computational capacities of modern computer hardware. 

The computational demands of deep learning approaches can however still represent a barrier to their practical use, e.g., because of the need for testing and optimizing different (hyper-)parameter configurations. Despite these limitations, researchers should not stop exploring more advanced algorithmic approaches in the future, given the sometimes low quality impression of today's recommender systems in practice \cite{JannachResnickEtAl2016} and the fact that some of today's first deep learning based approaches are sometimes not yet much better than computationally more simple approaches \cite{JannachLudewig2017RecSys}.
More research in this area is also particular important as naive methods can have certain limitations like a bias to recommend mostly popular items, which are not captured by today's problem formulations and common performance measures like precision and recall.


\subsubsection{Distributed Item Representations}
\label{sec:embeddings}

\paragraph{Methods}
Distributed item representations are dense, lower-dimensional representations of the items. The representations are derived from sequences of events and preserve the sequential relationships between the items. Similar to latent factor models, every item is associated with a real-valued \textit{embedding} vector, which represents its projection into a lower-dimensional space in which certain item transition properties are preserved.
For example, some representations are based on the co-occurrence of items in similar contexts \cite{grbovic15prod2vec,vasile16meta}. Others translate pairwise transition probabilities into distances in a Euclidean space \cite{chen12embedding,chen13multispace}.
In \seq{} recommendation problems, we can recommend the next item(s) given the user's most recent actions by traversing the embedding space in a stochastic fashion \cite{chen12embedding} or by searching the nearest neighbors to the last item(s) that were explored by the user \cite{grbovic15prod2vec}.

\paragraph{\applicationexamples}
Distributed Item Representations have been explored, for example, in the domain of playlist generation.
\citet{zheleva2010www} build a session-level long-term Latent Dirichlet Allocation (LDA) model based on the community of users and merge it with a short-term LDA model based on the current user session.
\citet{chen12embedding} propose the Latent Markov Embedding (LME) approach, a regularized maximum-likelihood embedding of Markov Chains in the Euclidean space.
In their method, every item is projected into a space in a way that the distance between any pair of items in this space is proportional to their transition probability in a first-order Markov Chain. The learning procedure directly exploits the \textit{weak ordering} between tracks in playlists.
The resulting vector spaces can then be traversed stochastically to generate \textit{new} sequences (in their case, playlists) or \textit{continuations} of existing ones.
Later on, \citet{chen13multispace} extend LME by clustering items and by adding cluster-level embeddings to account for locality in item transitions. \citet{wu13karaoke}, on the other hand, propose a Personalized LME version where both items and users are projected into a Euclidean space in a way that the strength of their relationship is reflected by the projection.
This allows them to incorporate long-term user preferences into the model and to use them for recommendation. More recently, \citep{reddy16learning} employ a similar approach for the recommendation of learning courses. Students, lessons and assessments are embedded in a common latent skill space, in which the evolution of the student knowledge is formalized by a Hidden Markov Model.
The goal of \citet{feng15personalized} is personalized POI recommendation based on the \textit{last-$N$} points of interest visited by the user.
In their approach, they use a pairwise ranking function similar to BPR \citep{rendle09bpr} to condition the transition probabilities on the personalized ranking preferences of the users.

A different technical approach -- in their case to deliver personalized product ads -- was chosen by \citet{grbovic15prod2vec}. Their work is based on leveraging the so-called ``distributional hypothesis'', which in the domain of linguistics states that semantically equivalent words occur frequently in the same contexts\footnote{The distributional hypothesis forms the basis of the recent Word2Vec \cite{mikolov2013distributed} and GloVe \citep{pennington14glove} approaches for Natural Language Processing.}, for \seq{} recommendation. The proposed Prod2Vec recommendation method learns distributed item representations from sequences of emails containing purchase receipts. Similar to the LME method, every item is projected into a lower-dimensional space. Specifically, Prod2Vec uses the skip-gram model, 
which projects items that tend to have similar neighboring items (i.e., items that tend to be ``surrounded'' by the same set of items) close to each other.
Given the sequence of the last few 
observations for a user, the most similar items in the lower-dimensional space represent the recommendation candidates.
Different enhancements of the recommendation scheme of Prod2Vec were proposed, including the usage of time decay factors for older email receipts or the consideration of directed order dependencies between the interactions \citep{djuric14hidden}. Furthermore, \citet{vasile16meta} propose Meta-Prod2Vec, an enhanced version of the skip-gram model that conditions the item embeddings also on their metadata. \citet{greenstein-messica17session} employ Word2Vec and Glove to create item embeddings based on click sequences and item metadata, and use these embeddings to enhance session-based recommendation with RNNs.

A technically different approach to use distributed item representations to learn a model from session-based user data is proposed by \citet{tagami15modeling}. The authors use the Paragraph Vector (PV-DV) model \citep{le14distributed}, which learns an additional user vector along with the item representations. In this approach, the user vectors computed from one day of interaction data are used as features in a multi-label classification problem to predict the set of ads the user will click on the next day. 

\paragraph{Discussion}
Distributed representations were successfully applied in particular in the domain of Natural Language Processing \cite{mikolov2013distributed} and, in the context of recommender systems, as an alternative method for representing textual data in content-based approaches \cite{almahairi15learning}. 
While the applications discussed in this section indicate that using such representations can be helpful for \seq recommendation scenarios as well, some embedding approaches can be computationally demanding and sometimes require extensive parameter tuning to yield good results. Similar to sequence modeling methods, approaches based on distributed representations require a substantial amount of training data to be effective.

While embedding methods exploit the sequential relations between items to learn the item representations, these methods do not make direct use of the sequence of recent user interactions to generate the recommendations. They instead rely on approximate methods like time-decay nearest neighbors \cite{grbovic15prod2vec} or on additional supervised learning layers \cite{tagami15modeling,baezayates15next} to predict the next interactions of the user. This in turn can lead to limited effectiveness in domains with strict ordering constraints for which sequence modeling methods can be preferable. Using an indirect approach can also further increase the computational costs of these methods.

For some works, the significance and practicability of some of the proposals in real-world environments is not always fully clear. The playlist generation method of \cite{chen12embedding}, for example, is not only computationally very intensive, it was also evaluated with a specific measure (the average log likelihood), which does not truly inform us about the quality of the resulting playlists \cite{bonnin14playlists}.


\subsubsection{Supervised Learning with Sliding Windows}
\begin{figure*}[t]
    \centering
    \includegraphics[width=0.6\textwidth,clip]{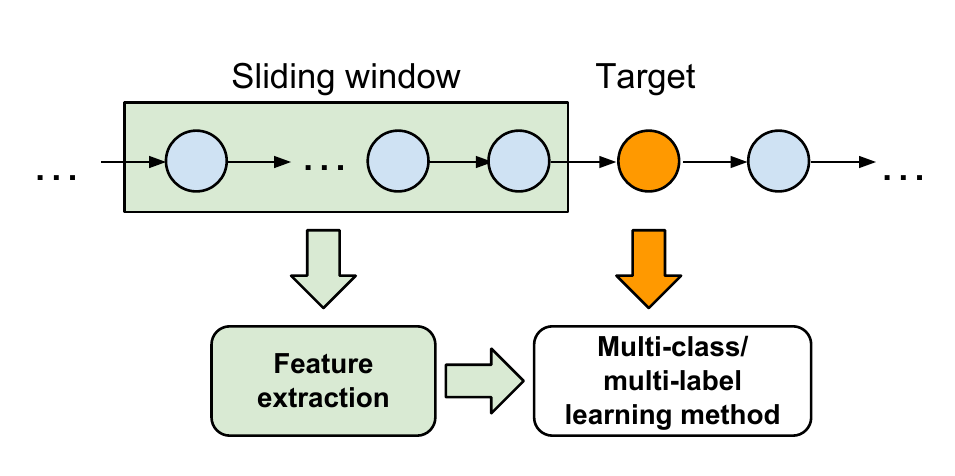}
    \caption{\Seq{} recommendation as supervised learning with sliding windows.}
    \label{fig:supervised}
\end{figure*}

\paragraph{Methods}
Sliding window models convert the next-in-sequence prediction problem into a traditional supervised learning problem that can be solved with any classifier such as decision trees, feed-forward neural networks and learning-to-rank methods. The general idea of the approach, which resembles autoregressive models, is as follows. 
A sliding window of size $W$ is moved over each sequence (see \autoref{fig:supervised}). At each step, all items within the window are used to derive the feature values of the supervised learning problem and the identifier of the immediately next item is used as target variable. As a result, the sequence prediction problem is turned into a multi-class classification problem, or into a multi-label classification in case multiple target items are allowed.

\paragraph{\applicationexamples}
In an early approach of that type, \citet{zidmars01temporal} frame sequential click prediction in a web usage mining scenario as a binary prediction problem. In their approach, clickstream data is first expanded by defining a set of ``accumulator'' variables (\textit{lagged} and \textit{cache} variables) to represent the contextual and historical activity of the user. Then, a page-level probabilistic decision-tree model is trained and at recommendation time the pages are ranked according to the predicted probability of being the next page.

More recently, \citet{baezayates15next} use contextual features that were extracted from the last 12 hours of the app usage log to predict which (mobile) app will be used \textit{next} in a \textit{session-aware} scenario.
Their contextual features include some \textit{basic} features, such as the geolocation and phone usage characteristics, as well as \textit{session} features, which are basically the Word2Vec representations \citep{mikolov2013distributed} of the actions of the user in the sliding window. Finally, they use a parallelized version of the Tree Augmented Naive Bayes algorithm for the classification (prediction) of the next used app.

In the e-commerce domain, \citet{wang15basket} use Feed-forward Neural Networks for \textit{next-basket recommendation}.
For the predictions, only the items in the previous basket were used to rank the items for the next one (i.e., the window size is 1). The first layer of the neural network encodes the previous basket as fixed-size real-valued feature vector that is computed by taking the element-wise maximum (\textit{max-pooling}) or average (\textit{mean-pooling}) of the embeddings of the items of the basket. The second and final layer of the network learns to rank the items in the next basket by taking as input the concatenation of the user's current and previous basket's embeddings.

\paragraph{Discussion}
The main advantage of this class of approaches is that it is general, conceptually simple, and that a variety of existing classification methods and libraries can in many cases be applied off-the-shelf. There are, however, also a number of shortcomings. First, as in many classification problems, the effectiveness of the approach depends on the quality of the feature engineering phase. Finding the best features can be challenging and often the features are domain-dependent and cannot be re-used across domains. Second, the obtained results can be highly sensitive on the choice of the window size. Finally, setting up a multi-class (or multi-label) classification problem can be computationally expensive when the set of items grows. Furthermore, the quality of the learned model can be strongly affected by unbalanced distributions of the target variable, which is a common situation in real-life recommendation scenarios.

\subsection{\Seq Matrix Factorization}
\paragraph{Method}
\Seq recommendation is different in various ways from the traditional matrix-completion problem formulation. However, there are a few cases where sequence information, which is usually derived from timestamps, is also considered within algorithms that are designed for matrix completion.\footnote{In this section we focus on \emph{\seq} and time-interval-based algorithms and do not discuss general \emph{time-aware} collaborative filtering techniques as described, e.g., in \cite{campos14time} and \cite{koren09cf}.}

\paragraph{\applicationexamples}
\citet{zhao12intervals,zhao14clusters} focus on an e-commerce recommendation scenario, where the problem is to fill the missing cells of a given user-item purchase matrix.
Specifically, they study the problem of finding the \textit{timing for repeated recommendations}.
Their proposed method factorizes the matrix using a weighted loss function that maximizes the expected utility for the user. The novel aspect of their approach is that the utility is determined based on the observed time intervals between purchases of each pair of items. The resulting model can predict the personalized relevance of an item depending on the time at which the recommendations are generated and thereby takes the available sequence information into account.

\citet{yu12sequential} analyze the problem of generating interactive personalized story \textit{continuations}.
Stories are represented in a prefix graph, in which each node represents a prefix of a story, i.e., a possible sequence of point plots. In their problem encoding, they are given a ``prefix-rating matrix'' to be completed, where the items are replaced by possible story prefixes and users provide ratings only to some of the prefixes. Matrix factorization is used to predict the missing entries. 
The highest scoring full story that descends from the current ``story-so-far'' of the user is used to suggest the next plot point. The rating of the user is collected on the next plot point, and the process (factorization, recommendation, rating) continues iteratively  until the story reaches an end.

Finally, \citet{twardowski16contextual} not only propose an RNN-based method as discussed above but also a matrix factorization approach to \textit{session-based recommendation} for the next-in-sequence recommendation problem in e-commerce.
In this approach, session events such as click, add-to-cart, etc., as well as the items are encoded by separate latent feature vectors. Sessions are then represented by the time-decayed sum of event vectors associated with the actions performed by the user so far in the current session. The prediction of the next items is then based on the factorization of the observed session-item tuples, which can be obtained with standard ranking loss functions.

\paragraph{Discussion}
The advantage of the discussed approaches is that standard matrix factorization algorithms from the literature can often be applied. A main challenge however lies in the definition of a suitable and computationally feasible encoding of the given application problem. Considering, for example, the proposal in \cite{yu12sequential}, one not only has to collect user feedback at each step of the sequence, but the given encoding can easily lead to a huge matrix due to the combinatorial explosion of the possible sequences. It also requires continuous updates of the factorization model as users have to provide new ratings for every suggested next step in the sequence.

The approaches discussed in this section can be seen as special cases of matrix factorization algorithms for time-aware recommendation (e.g. \cite{koren09cf}). These algorithms mostly focus on tracking changes of user behavior over large time spans, usually through time-evolving user and item latent factors.
Such methods, as discussed in more detail on \cite{campos14time}, can however often not be updated in real-time. In some application domains they should therefore be extended or combined with methods that support the short-term adaptation according to the user's current preferences \cite{jannach15adaptation}.

\subsection{Hybrid methods}
\paragraph{Methods}
Hybrid models often combine the flexibility of sequence learning methods with the robustness to data sparsity of factorization-based matrix-completion techniques. Furthermore, such forms of hybridization enable sequence learning methods to use the power of modern collaborative learning-to-rank models such as BPR \cite{rendle09bpr}, which usually cannot be easily embedded in standard sequence learning approaches.\footnote{The works presented in \cite{hidasi_session-based_2016} and \cite{hidasi16feature} represent exceptions that combine RNNs with BPR's ranking loss criterion.}

\paragraph{\applicationexamples}
Among the first proposals for such a hybrid technique is the Factorized Personalized Markov Chain (FPMC) method of \citet{rendle10FPMC}. The method combines matrix factorization with Markov Chains for the problem of \textit{next-item recommendation} given the \textit{last-$N$ interactions} of the user, e.g., in e-commerce settings.
In the case of first-order MCs, user interactions can be represented as a three-dimensional (\textit{user}, \textit{current item}, \textit{next item}) tensor. Each entry in the tensor corresponds to an observed transition between two items performed by a specific user.
The proposed method then uses pairwise factorization to predict the unobserved entries in the sparse tensor, i.e., to predict personalized transitions between pairs of items. Overall, FPMC can be seen as a first-order Markov Chain whose transition matrix is jointly factorized with a standard 2-dimensional user-item matrix factorization approach. This joint factorization at the end makes it possible to infer the unobserved transitions in the Markov Chain from the transition pairs of other users. At recommendation time, the items are ranked according to their likelihood to be the next item given the last item the user has interacted with.

In addition to e-commerce settings, the FPMC method has been successfully applied to other problem settings, e.g., for POI recommendation or check-in prediction in location-based social networks (\cite{he16inferring,lian13checkin}). \citet{he16fusing} present a variation of the FPMC method where the matrix factorization technique is replaced by a factored Item Similarity Model (FISM) \citep{kabbur13fism}. The resulting method, named Factorized Sequential Prediction with Item Similarity Model (Fossil), models users as a combination of the factors of the items they have interacted with, which in turn allows Fossil to provide sequential recommendations also for cold-start users -- the FPMC method is limited in that respect since it relies on user-item matrix factorization -- as long as the item representations can be estimated accurately. The method was tested in different application domains such as clothes, toy, or electronic devices recommendation. 

A number of other hybrid approaches were proposed in the literature. In the context of playlist recommendation, \citet{hariri12context} use Latent Dirichlet Allocation (LDA) \citep{blei03lda} to extract latent topics from playlists, where playlists are taken as documents and tracks as words for the LDA step. A sequential pattern mining technique is then applied to find patterns of such latent topics in the playlists.
When generating \textit{continuations} from existing sessions or playlists, the frequent patterns are mapped to the topics extracted from the current listening session and used to filter the recommendations that are generated by a classical nearest-neighbor-based recommender.

\citet{song15next} propose the States Transition Pair-Wise Ranking Model, which combines LDA with first-order MCs to simultaneously model the user's long and short-term favorites. The user's long-term favorites, e.g., in the movie domain, are determined by the topics generated in the LDA step with a user-specific prior.
The short term favorites are captured by an MC transition matrix. The combined model is then basically a HMM whose latent states are controlled by the personalized LDA generative model, which can be trained via Markov Chain Monte Carlo (MCMC) Bayesian inference and then be used to predict the \textit{next-item} given the last few interactions of the user.

Finally, in the context of next-app recommendation, \citet{natarajan13app} propose a method based on behavioral clustering to introduce personalization into \textit{session-aware} models.
In the behavioral clustering step, users with similar sequential
behavior are grouped by applying \textit{k-means} clustering on the per-user transition matrices of a first-order MC. A personalized PageRank algorithm is then used to build transition models at the cluster level. At recommendation time, the user is mapped to its corresponding cluster and the cluster-level transition model is used to provide sequential recommendations.

\paragraph{Discussion}
Hybrid approaches are often used to build recommendation systems due to their capability of
overcoming the shortcomings of individual methods, e.g., in terms of limited content discovery support or in the context of user- or item cold start situations \cite{burke07hybrid}.
The typical challenges when designing hybrid recommender systems include the problem of finding the best way of combining the predictions of the different recommendation channels and of determining importance weights for each individual method in the final prediction.

Some of the discussed approaches circumvent these issues by combining different competing methods in a unified model, like the combinations of Markov Chains with matrix factorization \cite{rendle10FPMC} or with LDA \cite{song15next}, in a way that the final recommendations are computed by a single algorithm. Other approaches rely on a cascade of algorithms \cite{hariri12context}, where a sequential model is used to filter the predictions generated by a sequence-agnostic one, or a meta-level approach \cite{natarajan13app}, where a first \seq model is used as basis for deriving another one.

\subsection{Other methods}
Only a handful of the methods proposed in the literature do not fit into the above categories, and most of them are either \emph{graph-based} or based on \emph{discrete optimization} techniques.


\paragraph{Graph-based methods}

\citet{xiang10temporal} focus on \emph{context adaption} to detect short-term user intents and interests.
The paper describes a graph-based approach which is evaluated based on implicit feedback datasets in the area of social bookmarking.
In their approach, long-term and short-term user preferences are fused into a two-sided bipartite graph, the Session-based Temporal Graph. One side of the graph connects users with the items they have interacted with in the past. The other side of the graph connects session identifiers with the items the user has interacted with in the current session. The edges are weighted to balance the influence of long-term and short-term preferences. The relationships between the items are propagated through the graph via Injected Preference Fusion, and at recommendation time the graph is traversed via a random walk to generate session-aware recommendations.

\citet{liu16unified} aim to detect \emph{interest trends} of individual users over time for POI recommendation.
Their method integrates static user interests and evolving sequential preferences based on temporal interval assessments.
Similar to \cite{zhao12intervals}, the time intervals between POI visits are assessed from user check-in sequences in a POI-to-POI transition matrix. Then, a bi-weighted low-rank graph is constructed to learn the individual user's behavioral preferences by identifying a set of common graph bases. As a result, the static interests and evolving sequential preferences of the user are learned simultaneously with the graph. As in \cite{zhao12intervals}, recommendations are generated by ranking the POIs by their expected relevance for a given time period.

In the same application domain, \citet{zhang14lore} try to detect if there are \emph{order constraints} when visiting POIs.
Their method, named LORE, mines sequential patterns from location sequences and represents the patterns as a location-location transition graph. The probability of a user visiting a new location 
is modeled through an additive $n$-th order Markov Chain, in order to incorporate sequential dependencies between locations.

Finally, \citet{trevisiol14cold} apply \emph{contextual adaptation} to address the \emph{new user} problem in news recommendation.
In their approach, the users' browsing behavior is represented by two graphs: the BrowseGraph, which collects the behavior of all user sessions, and the ReferrerGraph, a subgraph of the BrowseGraph which is induced by browsing sessions of users coming from the same referrer domain\footnote{Examples of referrer domains considered in the paper are \textit{Facebook}, \textit{Twitter} and \textit{Reddit}.}. When the user enters the news portal from the referrer domain, the recommendation candidates are chosen among the neighbors of the article in the Referrer Graph, using the Browse Graph as fallback. The article's neighbors are ranked according to various strategies, such as by random, by popularity, by content similarity and by edge weight.

\paragraph{Discrete optimization methods}
Discrete optimization is often employed for \seq recommendation problems when weak or strict ordering constraints between items exist. Typical application examples include travel planning, learning course sequence generation, and playlist generation.

Both \citet{jannach15playlist} and \citet{pauws06fast} address the \emph{list continuation} problem to generate playlists in the music domain.
\citet{jannach15playlist} propose a two-stage method to determine suitable continuations to a music listening session. In a first step, the recommendable tracks are scored based on a variety of features (including track co-occurrences and musical characteristics). Then, the first elements of the resulting list are re-ranked in a greedy approach in order to optimize the coherence of the track continuations with the recently listened tracks.
\citet{pauws06fast} include explicit \emph{order constraints}, either global to all users or local to a single user, in their model.
The authors first formulate the problem as an integer linear programming (ILP) problem, which is NP-hard in its solution, and then implement a local search heuristic to make the solution scalable.

Similar to \cite{pauws06fast}, \citet{xu16personalized} address the \emph{list continuation} problem with explicit \emph{order constraints}, but in a different application domain: personalized course sequence recommendation.
The specific goal is to find a sequence of courses that (a) will  minimize the time-to-graduation of the students and maximizes their GPA, (b) at the same time matches the interests of the user and (c) respects the sequential constraints between the courses (called the prerequisite graph). Similar to \cite{jannach15playlist}, they propose a two-stage algorithm. First, candidate sequences  with short time-to-graduation are extracted from the prerequisite graph using a Forward-Search Backward-Induction algorithm. Then, an online regret minimization algorithm based on multi-armed bandits is used to select course sequences that are expected to maximize the students' GPAs.


Generally, most discrete optimization methods for \seq recommendation do not rely on exact or exhaustive search. 
Instead, due to the computational complexity of the underlying problems, they usually resort to heuristic search or greedy optimization techniques, see also \cite{JugovacJannachLerche2017eswa}.


\subsection{Summary and Pros and Cons of Selected Approaches}
The main ideas of the most important families of algorithms along with a selection of their most typical advantages and disadvantages are summarized in \autoref{tab:algo_summary}. Note that the entries in the table mainly serve as a rough orientation and that specific pros and cons for individual algorithms within each family can exist.


\begin{table}
\caption{Summary of the Main ideas, Pros and Cons of the algorithms for \seq recommendation.}
\label{tab:algo_summary}
\setlength{\tabcolsep}{4pt}
\footnotesize
\centering
    \begin{tabular}{@{}p{1.3cm}p{4cm}p{4cm}p{4cm}@{}}
    \toprule
    \textbf{Algorithm}                                     & \textbf{Main idea}                                                                                                                    & \textbf{Pros}                                                                                                    & \textbf{Cons}                                                                                                                                           \\ \midrule
FPM                 & Discover patterns in user action sequences                                                                                   &
\begin{tabular}[t]{@{\textbullet~}p{4cm}@{}} Easy implementation \\ Explainable results\end{tabular} &
\begin{tabular}[t]{@{\textbullet~}p{4cm}@{}} Complex configuration \\ Suffers from data sparsity \\ Limited scalability \end{tabular} \\ \midrule

MC                            & Compute transition probabilities over fixed-length sequences                                                                & \begin{tabular}[t]{@{\textbullet~}p{4cm}@{}} Explainable results \end{tabular} & \begin{tabular}[t]{@{\textbullet~}p{4cm}@{}} Fixed transition order \\ Suffers from data sparsity \\ Limited scalability    \end{tabular} \\\midrule

VMM            & Compute transition probabilities over variable-length sequences                                                              & \begin{tabular}[t]{@{\textbullet~}p{4cm}@{}}  Variable transition orders \\ Explainable results                                        \end{tabular} &  \begin{tabular}[t]{@{\textbullet~}p{4cm}@{}} Suffers from data sparsity \end{tabular} \\\midrule

HMM                    & Model the causal factors in user sequences as transitions between \textit{discrete} hidden states                                                 & \begin{tabular}[t]{@{\textbullet~}p{4cm}@{}}  Learns from variable-length inputs \\ Robust to data sparsity   \end{tabular}         & \begin{tabular}[t]{@{\textbullet~}p{4cm}@{}}  Limited explainability \\ Huge number of discrete parameters \end{tabular}                                                   \\ \midrule

RL                   & Directly maximize the customer and seller reward over time                                                               & \begin{tabular}[t]{@{\textbullet~}p{4cm}@{}}  Dynamically adapt recommendations to future (unknown) rewards \\ Under active research  \end{tabular}                  & \begin{tabular}[t]{@{\textbullet~}p{4cm}@{}}  MDP-based approaches have same issues as MCs \\ Limited explainability \end{tabular}                                                         \\ \midrule

 RNN               & Model the causal factors in user sequences with \textit{non-linear} transitions between \textit{continuous} hidden states                                                   & \begin{tabular}[t]{@{\textbullet~}p{4cm}@{}}  Learns from variable-length inputs \\ Learns long-term dependencies \\ Robust to data sparsity \\ Compact hidden states \\ Under active research \end{tabular} & \begin{tabular}[t]{@{\textbullet~}p{4cm}@{}}  Complex configuration \\ Limited explainability \\ Benefits not fully clear in some domains \end{tabular}                                                       \\ \midrule

EMB        & Embed items into latent spaces that preserves sequential transition properties                       &  \begin{tabular}[t]{@{\textbullet~}p{4cm}@{}}  Robust to data sparsity \\ Visually interpetable embeddings \\ Under active research \end{tabular}                                       & \begin{tabular}[t]{@{\textbullet~}p{4cm}@{}}  Need auxiliary methods to make recommendations \\ Limited explainability \end{tabular}                 \\\midrule
SL & Use supervised learning over features extracted from fixed-size sliding windows over sequences & \begin{tabular}[t]{@{\textbullet~}p{4cm}@{}}  Easy implementation \\ Use off-the-shelf supervised algorithms   \end{tabular}                       &  \begin{tabular}[t]{@{\textbullet~}p{4cm}@{}}  Explainability depends on the chosen supervised method  \\ Feature engineering \end{tabular}                                                                                    \\\midrule
MF & Define new inputs and loss functions for MF to handle sequences   & \begin{tabular}[t]{@{\textbullet~}p{4cm}@{}}  Extensive literature available \\ Robust to data sparsity \end{tabular} & \begin{tabular}[t]{@{\textbullet~}p{4cm}@{}}  Non-trivial input and loss design \\ Concerns regarding scalability \end{tabular} \\\midrule
\multicolumn{4}{@{}p{\linewidth}@{}} {\footnotesize \textbf{Algorithm:} FPM: Frequent Pattern Mining, MC: Markov Chains, VMM: Variable-order Markov Models, HMM: Hidden Markov Models, RL: Reinforcement Learning, RNN: Recurrent Neural Networks, EMB: Distributed Item Representations, SL: Supervised Learning w/ Sliding Windows, MF: Matrix Factorization}
    \\\bottomrule
    \end{tabular}
\end{table} 
\section{Evaluation of \seq recommender systems}
\label{sec:evaluation}

\subsection{Common Evaluation Approaches for Recommender Systems}

In academic environments, the evaluation of recommender systems is dominated by ``offline'' (simulation-based) experiments on historical rating or implicit feedback datasets. 
The common offline evaluation methodology when using the matrix completion abstraction is to split the given preference data into training and test splits, use the training data to learn a model and predict the held-out preferences based on this model. 
The quality of the outputs of an algorithm can then be assessed with the help of 
measures such as RMSE, or Precision and Recall. In addition 
one can analyze a number of other quality factors of the recommendations, e.g., in terms of the diversity of the list, the general popularity of the recommended items, or the algorithm's catalog coverage. 

User (laboratory) studies are an alternative to offline experiments. Such studies are often used to assess the potential impact of recommendations on the behavioral intentions of users, an aspect which cannot be determined based on simulation studies. Finally, field studies (e.g., in the form of A/B tests) are used to analyze the effects of recommenders on their users in real-world environments. This latter form of evaluation is however comparably rare in academic environments.

\subsection{Offline Evaluation of \Seq Recommenders}
While in principle all three mentioned evaluation forms (offline studies, user studies, field tests) can be applied for \seq recommenders, slightly different evaluation methodologies are used in offline studies. Generally, several aspects of the evaluation protocols used for \seq recommenders are closely related to those that are used for Time-Aware Recommender Systems (TARS), for which \citet{campos14time} provide a formalization and a detailed analysis. In our subsequent analysis we will therefore take this framework as a reference.

\subsubsection{Evaluation Methodologies} \citet{campos14time} discuss evaluation methodologies for TARS along three dimensions: data partitioning, definition of the target items, and cross-validation.

\paragraph{Dataset partitioning} In standard matrix completion setups, the partitioning of the ratings into training and test sets is often done by sampling ratings randomly. Common variants are to either take samples from the entire dataset or sample ratings per user. For TARS, one can additionally consider a \emph{rating order} criterion and, e.g., select the most recent ratings (of the individual user or the community) for the test set.
For \seq recommenders, the rating order criterion, i.e., in our case the sequence of actions, is already given by definition.

Two forms of splitting the data into training and test set are possible: \emph{event-level} and \emph{session-level} splitting.
When splitting the data at the \emph{event-level} we can assign all events before a certain point in time to the training set and the remaining events to the test set, as done in \citep{grbovic15prod2vec}. With this procedure, an event is therefore assigned to one of the sets independent of the user or the session it might belong to.
Similarly, one can apply a time-based splitting criterion at the \emph{session-level}, i.e., we do not split up sessions but consider the timestamp of the first event of a session to decide if the session goes in the training or test set \cite{turrin15large,hidasi_session-based_2016}.


Both event-level and session-level partitioning can be applied to either the whole set of users (\emph{community-level} partitioning) or to a subset of the users (\emph{user-level} partitioning).
In the latter case one can select a number of \emph{training users} and put all their data into the training set.
Interactions of the remaining \emph{test users} are further split -- either at event or session-level -- into a \emph{user profile} and \emph{test data} \citep{jannach15adaptation}.
The user profile includes the less recent interactions and is used as input to the recommender; the test data contains the most recent interactions to be predicted.
Finally, mixed approaches are in principle possible as well.
One can, for example, select a number of users as test users and put all data into the training set except for the most recent sessions of the test users \citep{quadrana17personalizing}.

Besides the question of the splitting criterion, different options exist of how to apply \emph{size conditions} \cite{campos14time}. 
Placing 80\% into the training set and 20\% in the test set 
is a common approach.
In \seq recommenders, additional options exist.
One can use \textit{time-based} splits and place events (or sessions) prior to a give time into the training set and use the rest for testing \cite{feng15personalized}.
Or, we can use \textit{fixed-size} splits and place the last $k$ events of each sequence into the test set and the remaining into the training set \citep{cheng13where}.


Our survey shows that no common standard exists in the community regarding data partitioning procedures.
In the context of session-based or session-aware recommendation problems, it is however advisable to use a session-level partitioning procedure to avoid that individual user sessions are split up.
This is in particular necessary to mimic the behavior of session-based and session-aware recommender systems that are batch-trained on instances of past sessions for efficiency reasons. 
In the specific case of session-aware recommendation problems, community-level partitioning has to be applied to ensure that a recommender can also learn longer-term models for individual users.
In the case of session-based recommender systems, user-level partitioning has to be preferred to ensure that a recommender is able to provide recommendations to new users (i.e., users with no records of past interactions with the system).

Also with respect to the size conditions, no community standards exist so far.
In some works, for example in \cite{hidasi_session-based_2016}, all sessions except those of the last day are placed into the training set.
While the maximum amount of data is provided for learning in such a setting, only one single training-test split can be used.
The size of the training data can also be chosen based on domain-specific requirements.
In domains such as news recommendation or e-commerce, focusing on the most recent interactions in some cases is sufficient or even advisable, e.g., because news can quickly become outdated or because e-commerce shoppers might concentrate on trending or recently added items \cite{JannachLudewig2017RecSys}.
In general, when using time-based splitting, the evaluation should be performed with different time splits, which allows us to evaluate the learning rate of the algorithm (i.e., the minimum amount of training data that provide stable recommendations).


\paragraph{Definition of the target items}
In traditional evaluation setups, we aim to predict ratings for a set of target items or search for an optimal ranking of these items. In \seq recommenders, we are usually interested to predict future \emph{actions} of a user (or, generally, events), where actions as described above usually have an associated type and item. In addition, the input to the evaluation step is not limited to a user or user-item pair for which a ranked list or rating prediction is sought for, but the input can also include a sequence of events, e.g., an entire session from the test set in session-based or session-aware recommendation scenarios.
If such a sequence of events is given, the general idea in most evaluation approaches is to hide a subset or all of the events of the given sequence and predict the user's (next) actions. Different variations of the evaluation scheme can be found in the literature.

\begin{itemize}
\item \textit{Sequence-agnostic prediction}. In this case, the order of the hidden actions is not relevant and a recommendation is considered successful if it correctly predicts one or more of the hidden actions \citep{zhao12intervals,zhao14clusters}.
\item \textit{Given-N next-item prediction}. In this protocol variant, the first $N$ elements of the sequence are ``revealed'' to the algorithm and the task of the recommender is to predict the immediate next action. \citet{jannach15adaptation} propose to use one specific value for $N$ when evaluating.
In the domain of next-track music recommendation, often only the last element of a sequence is hidden (i.e., $N=1$) \citep{hariri12context}.
    In \citep{hidasi_session-based_2016,hidasi16feature,quadrana17personalizing}, an approach is taken where $N$ is incrementally increased.
\item \textit{Given-N next-item prediction with look-ahead}. This is a combination of the above protocols and used in \citep{grbovic15prod2vec, hidasi_session-based_2016}: the set of revealed events is continuously increased.
The order of the hidden actions (which form the look-ahead set) is not relevant.
\item \textit{Given-N next-item prediction with look-back}. In addition to the first $N$ items of the current sequence, the idea of this protocol is to also reveal a set of actions that happened \emph{before} the current sequence \cite{jannach15adaptation,Lerche2016Reminder,quadrana17personalizing}.
\end{itemize}

Besides the different ways of evaluating the predictions for a given sequence or user session, one can limit the evaluation to actions of certain types, e.g., to purchase actions, as done in \cite{jannach15adaptation} and in the context of the 2015 ACM RecSys challenge\footnote{\url{http://2015.recsyschallenge.com/}}.

The choice of the best evaluation protocol depends both on the specific research question and on the application domain.
When evaluating session-based recommender systems, \textit{given-N next-item prediction} (with or without look-ahead) usually is preferred. In case the goal is to assess how quickly a system is able to adapt its recommendations to the user's assumed short-term intents, one might use a \emph{given-N} protocol where $N$ is incrementally increased.
Hiding only the last element of a sequence
can be done in different application scenarios to assess the recommendation performance when more information is available (``warm start'').
Finally, when the goal is to evaluate the value of combining short-term and long-term user preferences or the effectiveness of different \emph{reminding} strategies, the \textit{given-N next-item prediction with look-back} protocol can represent an appropriate choice.



\paragraph{Cross validation} Using a randomized cross-validation procedure is not possible in all application areas of \seq recommenders. In the case of next-track music recommendations, where the goal is to predict the next track within a given playlist \cite{hariri12context,jannach15playlist}, the given sets of playlists can be randomly split into training and test sets as long as the creation time of the playlists is not relevant.

In many other cases, however, sequence or time dependencies exist, which make it impossible to randomly assign events to the training and test splits. Some works like \citep{hidasi_session-based_2016} therefore only use one single training-test split, 
which can however lead to biased results. Therefore, alternative validation approaches like repeated random subsampling procedures should be applied. One can, for example, split the data into several, potentially overlapping time slices of about the same length (``sliding window'') and use the events at the end of these time slices as test sets \cite{jannach15adaptation,JannachLudewig2017RecSys}. Alternatively, in case of large datasets, one can in addition repeatedly sample a subset of the users and consider their last $n$ actions in the test set.


\subsubsection{Evaluation Metrics}
For many application scenarios of \seq recommenders, standard \emph{classification} and \emph{ranking} metrics can be used for performance evaluation.
The set of metrics used in the reviewed papers include Precision, Recall, Mean Average Rank (MAR), Mean Average Precision (MAP), Mean Reciprocal Rank (MRR), Normalized Discounted Cumulative Gain (NDCG) and the F1 metric.
Typically, most of the ranking metric are highly correlated when evaluated with a fixed-size split and their choice does not largely affect the outcomes \cite{herlocker2004evaluating,cremonesi2010performance}.

For some application domains, such as context adaptation, we can use standard classification metrics (e.g., precision and recall) since the recommendation problem is a classical one and sequence-based techniques are used to generate better recommendations. However, when the application domain requires recommendations to fulfill an explicit or implicit order constraint, ranking metrics such as MAP have to be preferred over classification metrics.
\textit{Error} metrics were considered only in \citep{yu12sequential} and, in general, do not fit sequence-aware recommendation problems well, where in almost all situations it is important to consider entire lists of recommended items. 

In some application domains ranking metrics alone cannot fully inform us about the quality of the recommendations.
In playlist recommendation, for example, we may want to assess the capability of the recommender in generating good quality transitions between subsequent songs. Given a current track, there may be many other tracks that are almost equally likely good to be played next, and ranking metrics to not take that aspect into account.
This can in turn lead to the effect that more popular tracks are favored by an algorithm \cite{JannachLercheEtAl2015}. Therefore, alternative metrics derived from natural language processing, like the Average Log-Likelihood, can be adopted instead \citep{chen12embedding,chen13multispace}. However, several concerns on the actual quality of the recommendations and on the interpretability of the results with these metrics have been raised \cite{bonnin14playlists}.

Generally, when the goal of a recommender is not limited to predicting the next-best action but a sequence or set of events that are related to each other, alternative ``multi-metric'' evaluation approaches are required that can take multiple quality factors into account in parallel. They can consider, as mentioned, the order of the recommendations or the internal coherence or diversity of the recommended list as a whole. In the music domain, one might also be interested that the set of next tracks is coherent not only in itself but with the last played tracks. In many cases, the different quality factors can lead to trade-off situations like ``accuracy vs.~diversity'', which have to be balanced by a recommendation algorithm \cite{JugovacJannachLerche2017eswa}.

Differently from conventional recommender systems, many sequence-aware application domains recommend also items that the user already knows or has purchased in the past (see Section \ref{sec:repeated}). In these application scenarios, the evaluation protocol has to be adapted in order to consider already known items when computing the metrics.

Finally, sequence learning methods are often computationally more complex than traditional matrix-completion algorithms.
Therefore, the evaluation of sequence-aware recommender systems should also discuss the time and space complexity of the methods and report running times for typical training data sets so that the scalability can be assessed.

Overall, a general open issue in the field is the lack of ``standard'' metrics to assess quality criteria when recommendation lists as a whole should be evaluated, e.g., with respect to their diversity or coherence.
Usually, a selection of meta-data features is used to measure, for example, \emph{intra-list} diversity or the smoothness of the transitions between the objects in the list.
However, to which extent such measures reflect the users' quality perceptions often remains largely open.

\subsubsection{Data Sets}
In recent years, more and more datasets to benchmark different \seq algorithms in a reproducible way have become available to researchers. Table \ref{tab:datasets} shows the characteristics of a number of public datasets that were used in existing works.

Some researchers rely on traditional rating datasets with time stamps to evaluate \seq algorithms. In reality, however, the point in time when users rate an item might be quite different from the one when the user experienced the item (e.g., watched a movie). The design of corresponding algorithms and the interpretation of the results must therefore be done with care.


\begin{table}
\caption{Publicly available datasets for \seq recommendation.}
\label{tab:datasets}
\footnotesize
\begin{tabular}{@{}llllllP{4cm}@{}}
\toprule
\textbf{Short name}&\textbf{Domain}&\textbf{Users}&\textbf{Items}&\textbf{Events}&\textbf{Sessions}&\textbf{Reference}\\
\midrule
Amazon&EC&20M&6M&143M&N/A&\citep{he16fusing}\\
Epinions&EC&40k&110k&181k&N/A&\citep{he16fusing}\\
RecSys Chall. 2015&EC&N/A&38k&34M&9.5M&\citep{hidasi_session-based_2016}\\
Ta-feng&EC&32k&24k&829k&N/A&\citep{wang15basket,yu16dynamic}\\
TMall&EC&1k&10k&5k&N/A&\citep{wang15basket,yu16dynamic}\\
AVITO&EC&N/A&4k&767k&32k&\citep{twardowski16contextual}\\
Retailrocket&EC&1.4M&235k&2.7M&N/A&\citep{retailrocket_dataset}\\
\midrule
Microsoft&WWW&27k/13k&8k&13k/55k&32k/5k&\citep{zidmars01temporal,zhou04intelligent}\\
MSNBC&WWW&1.3M/87k&1k&476k/180k&N/A&\citep{zidmars01temporal,yap12sequential}\\
Delicious&WWW&8.8k&3.3k&60k&45k&\citep{xiang10temporal}\\
CiteULike&WWW&53k&1.8k&2.1M&40k&\citep{xiang10temporal}\\
Outbrain&WWW&700M&560$^\ast{}$&2B&N/A&\citep{outbrain_dataset}\\
\midrule
AOL&Query&650k&N/A&17M&2.5M&\citep{sordoni15hierarchical}\\
\midrule
Adressa&News&15k&923&2.7M&N/A&\citep{gulla17adressa}\\
\midrule
Foursquare 1&POI&31k&N/A&N/A&N/A&\citep{he16inferring}\\
Foursquare 2&POI&225k&N/A&22.5M&N/A&\citep{cheng13where,he16fusing}\\
Gowalla 1&POI&54k&367k&4M&N/A&\citep{cheng13where,he16fusing}\\
Gowalla 2&POI&196k&N/A&6.4M&N/A&\citep{liu16next,liu16unified}\\
NYC Taxi Dataset&POI&N/A&299&N/A&670k&\citep{du16recurrent}\\
Global Terrorism DB&POI&N/A&N/A&N/A&N/A&\citep{liu16next}\\
\midrule
Art Of the Mix&Music&N/A&218k&N/A&29k&\citep{hariri12context,mcfee11nlp,jannach15playlist}\\
30Music&Music&40k&5.6M&31M&2.7M&\citep{turrin15large,vasile16meta}\\\bottomrule
\multicolumn{7}{@{}p{\linewidth}@{}} {\scriptsize EC: E-commerce, WWW: Web browsing; POI: points of interest; $^\ast{}$number of sites.}
\end{tabular}
\end{table}

\subsection{On User Studies and Field Tests}
User studies, which for example aim to evaluate the subjective quality perception of users in \seq recommendation scenarios are comparably rare. In a recent study in the context of next-track music recommendation \cite{KamehkhoshJannach2017}, researchers compared the quality perception of different algorithms. They designed an online experiment where the main task for the 277 participants was to pick one of the recommendations that were provided by four different algorithms as the best continuation of a given playlist.\footnote{A similar study in the music domain was presented in \cite{barrington2009smarter}.} One of the main results of the study was that considering musical features not only helps to increase the homogeneity of the recommendations in terms of computational (offline) measures, but also to a better quality perception by users. At the same time, focusing on generally popular tracks, which are in many cases already known to the users, showed to be a comparably ``safe'' strategy, at least in the short term. While the study provided evidence that in this particular domain, the chosen offline measures can be suitable proxies for certain aspects, user studies also have their limitations, e.g., in terms of the often artificial setup.

The number of reports on \emph{field tests} for \seq recommendation scenarios is limited as well. The few papers that exist include \cite{Garcin:2014:OOE:2645710.2645745}, \cite{grbovic15prod2vec}, \cite{JannachLudewigLerche2017umuai}, \cite{moling12optimal} and \cite{shani05mdp}. In \cite{JannachLudewigLerche2017umuai}, the authors for example report the results of an A/B test, where one strategy was to simply remind users of previously seen items in a session-aware recommendation scenario. In that experiment it turned out that reminding users of recently seen items is not only leading to comparably high accuracy in offline experiments, but also to a certain business value in the real world. Differently from this work, the studies of \citet{Garcin:2014:OOE:2645710.2645745} in the news domain revealed that the results of an offline experiment were not indicative for the performance of algorithms when deployed in a real system. Specifically, a session-based algorithm showed to be much more effective than a popularity-based strategy that worked best in the offline experiments. Overall, finding good proxy measures to predict the success of a recommender based on offline simulations is similarly challenging for \seq{} approaches as for other types of recommendation problems.


\section{Summary and Future Directions}
\label{sec:conclusions}

\Seq recommendation -- in particular in the forms of session-based and session-aware recommendation -- is a highly relevant problem in practice. Researchers have developed a variety of algorithmic proposals over the past fifteen years, and with this survey, our goal is to categorize these approaches and to review the research practice in the field. Throughout the paper, we have identified a number of open research questions, among them the following.

\emph{Intent detection.} Context-adaptation is one major goal in session-aware recommendation, and the most crucial task here is to estimate the users' context and short-intents. However, these intents cannot always be reliably estimated from the first few interactions of a session. For example in the music streaming domain, is the user in the mood to discover something new or rather interested in consuming known things \cite{kapoor15explore}?  In e-commerce, is the user currently browsing the catalog to understand the range of options, or is she or he interested in re-inspecting a short list of candidates to purchase? To address these questions, future works could for example consider existing works in the field of \emph{query intent understanding} from the information retrieval field, using, e.g., external knowledge \cite{Hu:2009:UUQ:1526709.1526773}. In addition, research could build on existing theories from other fields like Marketing, where models like AIDA (Attention, Interest, Desire and Action) are used to describe the different phases of a consumer, from becoming aware of an offering to the actual purchase. As discussed in \cite{Song2017When}, there are also domains where the user's interest can change even within a session, for example, when a user has read a number of news stories of a certain category and then becomes satiated with the topic. Finally, in the context of intent detection, very little research exists on how to give users the opportunity to correct the system's assumption in case they are wrong. On Amazon.com, some basic mechanisms for user control exist, but it is unclear if they are broadly used by consumers \cite{JannachNaveedEtAl2016}.

\emph{Combining short-term and longer-term profiles.} To be able to assess the consumers' state in the decision making process, information about their behavior during recent sessions and probably also their longer-term behavior has to be taken into account. The most recent works in the field focus on the session-based recommendation problem, where this past information is not available. However, there are many domains where longer-term user profiles exist and more research is required to better leverage this information. Some works \cite{jannach15adaptation} show that while the short-term intentions should be predominant in the selection of the recommendations, considering longer-term behavioral patterns and preferences of the individual user (e.g., toward certain brands in an e-commerce scenario) can be important. Existing works like \cite{Billsus:2000:LAW:325737.325768} use a simple static weighting of long-term and short-term models or rely on on re-ranking the results of the long-term models based on assumed short-term intents \cite{jannach15adaptation,JugovacJannachLerche2017eswa}. Better integrated models as used, e.g., in \cite{quadrana17personalizing} are therefore still needed.

\emph{Leveraging additional data and general trends.} Generally, in the context of user profiling, researchers often rely on one or a few types of specific user interactions like item view events or check-ins at certain locations. In real-world applications, usually much richer types of information are available. Not only are there multiple additional actions related to certain items (e.g., add-to-wishlist, add-to-cart in the e-commerce domain), there are also other relevant user actions like search or category navigation which are not considered to a large extent in today's research. Also the information about how the users entered the site (e.g., through a search engine) can be a valuable piece of information to assess the user intent from the first few interactions.
Going beyond behavioral patterns of the individual, more research is also required in terms of detecting behavioral trends and interest shifts in the entire user community. In different application domains where item recency plays a role, including news, music and e-commerce, being able to detect and leverage short-term trends in a community is highly important. For the e-commerce domain, works like \cite{JannachLudewig2017RecSys} or \cite{JannachLudewigLerche2017umuai} show that focusing on items that were trending on the platform on the last few days can be crucial to further improve the recommendation quality.

\emph{Toward standardized and more comprehensive evaluations.} Finally, our review shows that today a variety of protocol variants and computational metrics are used to compare \seq algorithms, making it difficult to assess how the field progresses. Going beyond the subtle details of the evaluation protocols (e.g., of how to specifically determine the hold-out set), today's research in \seq recommenders is still largely focused on accuracy measures. While this is a known issue also for the matrix-completion formulation of the recommendation problem, in \seq recommendation, and in particular in session-aware recommendation, the usefulness of certain recommendations can largely depend on the contextual situation of the users, e.g., if they are exploring the item space, inspecting a smaller choice set, looking for complements to a recently inspected items etc. These aspects call for more purpose-oriented evaluation approaches \cite{JannachAdomavicius2016purpose,cremonesi2013user}, which take the users' specific contextual situation and goals into account. 

\bibliographystyle{abbrvnat}

\bibliography{bibliography-short}



\end{document}